\documentclass[useAMS,usenatbib]{mn2e}
\usepackage{graphicx}
\usepackage{epstopdf}
\usepackage[figuresright]{rotating}
\usepackage{subfigure,color}
\usepackage{longtable}
\usepackage{tabularx,multicol,lipsum}
\usepackage{xspace}
\newcommand{\ion}[2]{{\textrm{#1}}{\textrm{\sc #2}}}

\definecolor{pink}{rgb}{.9,.2,.5}  
\definecolor{purple}{rgb}{.5,.6,.7}

\title[Oxygen abundance discrepancy in  Seyfert 2s]{Chemical abundances of  Seyfert 2 AGNs -- III.  Reducing the oxygen abundance discrepancy}

\author[Dors et al.]
            {O.~L. Dors$^{1}$\thanks{E-mail:olidors@univap.br}, R. Maiolino$^{2,3}$,  
	    M.~V.~Cardaci$^{4,5}$, G.~F.~H\"{a}gele$^{4,5}$, A.~C. Krabbe$^{1}$, \newauthor{E.~ P\'erez-Montero$^{6}$, M. Armah$^{1}$} 
               \\
$^{1}$ Universidade do Vale do Para\'iba, Av. Shishima Hifumi, 2911, Cep
12244-000, S\~ao Jos\'e dos Campos, SP, Brazil\\ 
$^{2}$ Cavendish Laboratory, University of Cambridge, 19 J. J. Thomson Ave., Cambridge CB3 0HE, UK \\
$^{3}$ Kavli Institute for Cosmology, University of Cambridge, Madingley Road, Cambridge CB3 0HA, UK \\
$^{4}$ Instituto de Astrof\'{i}sica de La Plata (CONICET-UNLP), Argentina \\
$^{5}$ Facultad de Ciencias Astron\'{o}micas y Geof\'{i}sicas, Universidad Nacional de La Plata, Paseo del Bosque s/n, 1900 La Plata, Argentina \\
$^{6}$ Instituto de Astrof{\'i}sica de Andaluc{\'i}a, Camino Bajo de Hu{\'e}tor s/n, Aptdo. 3004, E18080-Granada, Spain.
} 
\begin{document}

\date{Accepted 2015 Month  00. Received 2015 Month 00; in original form 2014 December 17}

\pagerange{\pageref{firstpage}--\pageref{lastpage}} \pubyear{2011}

\maketitle

\label{firstpage}

\begin{abstract}

We investigate the discrepancy between oxygen abundance estimations for narrow-line regions (NLRs) of
 Active Galactic Nuclei (AGNs) type Seyfert 2 derived by using direct estimations of the electron temperature ($T_{\rm e}$-method) and 
those derived by using photoionization models.  In view of this, observational   emission-line ratios  in the 
optical range ($3000 \: < \: \lambda(\rm \AA) \: < 7000$) of Seyfert 2 nuclei 
compiled from the literature were reproduced by detailed photoionization models built with the {\sc Cloudy} code.
We find that the derived discrepancies are mainly due to the inappropriate
use of the relations between temperatures of the low ($t_{2}$) and high ($t_{3}$) ionization 
gas zones derived for \ion{H}{ii} regions in AGN chemical abundance studies. 
Using a photoionization model grid, we derived a new expression  for    $t_{2}$  as a function of  $t_{3}$  valid for Seyfert 2 nuclei.
The  use of this new expression in the AGN estimation of the O/H abundances  based on  $T_{\rm e}$-method   produces O/H abundances 
slightly lower (about 0.2 dex) than those derived from detailed photoionization models. 
We also find that the new formalism for the $T_{\rm e}$-method reduces by about 0.4 dex the O/H discrepancies
between the abundances obtained from strong emission-line calibrations and those derived from direct estimations.
\end{abstract}

%________________________________________________________________________

\section{Introduction}
\label{intro}

Active Galactic Nuclei (AGNs) and Star-forming regions (SFs) emit  strong metal emission-lines easily observable at practically  
all spectral ranges. The relative fluxes of these  emission-lines 
can be used to estimate their gas phase metallicity, among other properties, of these objects 
up to high redshifts. Therefore,  AGNs and SFs play a key role in studies of the chemical evolution of galaxies
across the Hubble time.

The relative abundance of oxygen to hydrogen (O/H) is usually used as a tracer of the total metallicity ($Z$) in galaxies 
since the prominent emission-lines from their main ionic stages are well detected in the
optical spectra in both SFs and AGNs (e.g.\ \citealt{alloin92, kennicutt03, hagele08, yates12}).
It is widely accepted that reliable  oxygen abundance  determinations  in gaseous nebulae
(i.e. \ion{H}{ii} regions, Planetary Nebulae) are those computed by direct estimations of the electron temperature, usually 
known as $T_{\rm e}$-method (see \citealt{enrique17, peimbert17, maiolino19} for a review).
Basically, this method consists of determining the electron
temperature ($T_{\rm e}$) of the gas phase through emission-line intensity ratios
emitted by a given ion  and originated  in transitions from two
levels with considerable different excitation energies, such as the 
$R_{\mathrm{O}3}$=[\ion{O}{iii}]($\lambda$4959+$\lambda$5007)/$\lambda$4363 ratio.
Although the first effort to discuss the chemical abundance in gaseous nebulae  was made by \citet{page36} and, later, by 
\citet{bowen39} and \citet{wyse42}, the first application of the
$T_{\rm e}$-method was carried out by \citet{aller54} for the Planetary
Nebula NGC\,7027 and by \citet{aller59} for  Orion nebulae.
 For AGNs, the first determination of abundance of heavy elements by using the $T_{\rm e}$-method
seems to have been carried out by \citet{osterbrock75}  for the radio galaxy 3C\,405 (Cygnus
A).  After this pionereeing work,  \citet{alloin92} applied the $T_{\rm e}$-method
for the Seyfert 2 galaxy ESO\,138\,G1  and \citet{izotov08} for AGNs located in four dwarf galaxies (see also \citealt{dors15, dors20}).
 Despite several other authors have addressed efforts  to
determine  chemical abundances in Narrow Line Regions (NLRs) of AGNs in the local universe (e.g. \citealt{ferland83, grazina84, ferland86,
cruz91, thaisa98, groves06, feltre16, castro17, enrique19, sarita20}) 
and at high redshifts (e.g. \citealt{nagao06a, matsuoka09, matsuoka18,  nakajima18, dors18, mignoli19, guo20}),  most
 studies have been based on  photoionization models.

The $T_{\rm e}$-method has been  used to compute
the abundance of heavy elements (e.g.\ O, N, S) for thousands of local \ion{H}{ii} regions and star forming galaxies
(e.g. \citealt{smith75, peimbert78, torrespeimber89, garnett97, vanzee98, kennicutt03, bresolin04, hagele06, hagele11, hagele12, zurita12, 
croxall16, lin17, fernandez18, esteban19, berg20}, among others)
and for some objects at high redshifts ($z \: > \: 1$; e.g.\ \citealt{sanders16, sanders20, gburek19}). 
Unfortunately, most of the objects for which the $T_{\rm e}$-method can be applied have high-excitation and low metallicity, as the
intensities of the required emission lines depend exponentially on the temperature of the gas.
For this reason, in many cases it is necessary to use calibrations of the total oxygen abundance with the relative fluxes of 
other detected strong lines (see e.g.\ \citealt{vanzee98}), as proposed by  \citet{jensen76} and \citet{pagel79}.  
The problem is that abundance values calculated through the  $T_{\rm e}$-method and those by using 
 calibrations from photoionization models are in disagreement with each other   by 0.1-0.4 dex (e.g. \citealt{kennicutt03, dors05, angel07, kewley08}).
However, there are other model-based abundances that do not present any discrepancy with the direct method
(e.g. \citealt{enrique10}, \citealt{enrique14}).
The origin of the  discrepancy  is an open problem in the nebular astrophysics and it can be
due to, for instance, the presence of electron temperature fluctuations in \ion{H}{ii} regions
\citep{peimbert67}, departure from  Maxwell-Boltzmann equilibrium energy
distribution, i.e. the fact that electron temperature in the gas may be better described by the
$\kappa$ distribution  \citep{nicholls12,binette12},  inappropriate use of Ionization Correction Factors (ICFs),
or uncertainties in photoionization models (e.g.\ \citealt{viegas02, kennicutt03})
generally used to obtain calibrations. 
However, all scenarios above do not provide a proper explanation for this discrepancy problem.

Regarding AGNs, the metallicity discrepancy problem is even more pronounced than in SFs.
For instance,  \citet{dors15} compared the total oxygen abundances obtained from the  $T_{\rm e}$-method
in a sample of Seyfert 2 galaxies with  other independent estimations including  
photoionization models and the extrapolations of the O/H gradient to the
nuclear region of the host spiral galaxies. They found that the $T_{\rm e}$-method  produces unrealistically low sub-solar abundance
values underestimating the oxygen abundances by up to $\sim 2$ dex  
(with an average value of $\sim 0.8$ dex) in relation to the other two methods.
Moreover, \citet{dors15} showed that  this
discrepancy is systematic, in the sense that it increases as the metallicity decreases.
Recently, this result was confirmed by \citet{dors20}, who used an homogeneous sample of  153 confirmed Seyfert 2 nuclei
from the Sloan Digital Sky Survey (SDSS, \citealt{abazajian09}) DR7.
This discrepancy is also known as the 'temperature problem'
and it has been attributed to the difficulty in reproducing high
electron temperatures such as those derived from the observational $R_{\mathrm{O}3}$ emission-line ratio with photoionization models,
which translates into an O/H abundance discrepancy. 

During the past years, several  authors have proposed some scenarios to explain
the high electron temperature estimated for the gas phase of AGNs, not reproduced by photoionization models.
\citet{heckman79} argued that electron temperature values higher than 20\,000 K require a secondary source of energy in addition 
to photoionization, possibly the presence of  shocks (e.g.\ \citealt{zhang13, contini17a}). 
However, signatures of the presence of very strong shocks are not found in the Seyfert 2 spectra since 
narrow (permitted and forbidden) emission-lines show typical line widths in the order of 100-600 $\rm km \: s^{-1}$ 
(e.g. \citealt{zhang13}). 
\citet{komossa97} built multi-component photoionization models considering
electron density inhomogeneities to interpret the observed narrow optical emission-line intensities of Seyfert 2 nuclei.
Even though these models reproduce the [\ion{S}{iii}]$\lambda$9069,$\lambda$9531 emission lines, as well as high-ionization
lines such as [\ion{Fe}{vii}]$\lambda$6087, they fail to reproduce the [\ion{O}{iii}]$\lambda$4363/$\lambda$5007 lines ratio.
Despite the $T_{\rm e}$-problem is, probably, the main cause of the O/H discrepancy in Seyfert 2, some additional
questions arise from the applicability of the $T_{\rm e}$-method in this kind of object. Some other authors
applied the $T_{\rm e}$-method to calculate the elemental abundances in AGNs but did not analyze its validity 
by comparing their results with those obtained using other methods. For example, \citet{izotov08}  
calculated the O/H abundance using the $T_{\rm e}$-method for AGNs in four dwarf galaxies and found very low abundances ($\rm 12+\log O/H$=7.36-7.99).
Although no alternative methods were considered by these authors, low O/H values are expected because the objects considered
are weak AGNs implying that the emission lines are not necessarily dominated by the
NLR emission. Moreover, these AGNs are located at low mass galaxies, which are expected to have low metallicity 
\citep{lequeux79, maiolino08, matsuoka18, thomas19}.  
 In any case, authors  who considered  $T_{\rm e}$-method to derive abundances in AGNs 
have assumed the formalism  developed for \ion{H}{ii} regions, which is not necessarily applicable to AGNs. 
In summary, ($i$) the origin of the O/H discrepancy is unclear in AGNs and ($ii$) it is ill-defined if it has the same origin as \ion{H}{ii} regions.
Within this context, it is necessary to explore the applicability of the $T_{\rm e}$-method for AGN studies.

In a previous paper, \citet{dors17} performed detailed photoionization
modelling to reproduce optical emission-line intensities  of a sample of   Seyfert 2  nuclei in order to derive the N and O abundances.
In this work, we used these detailed models and the same observational sample to investigate the O/H discrepancy origin in Seyfert~2 nuclei.
This paper is organized as follows: in Section~\ref{meth}
the methodology used in this paper is presented; in Section~\ref{res} we present the
comparison between the oxygen abundance  values derived using the $T_{\rm e}$-method and those from photoionization models, while discussion and conclusions
of the outcome are given in Sections~\ref{disc}  and \ref{conc}, respectively.

\begin{table*}\small
\caption{De-reddened fluxes (relative to H$\beta$=1.00)   for  a sample of Seyfert 2 nuclei.
The observed values compiled from the literature (see Sect.~\ref{obs}) are  referred as "Obs." while the   predicted values
by the photoionization models (see Sect.~\ref{model}) as  "Mod.".}
\label{tab1}
\begin{tabular}{lccccccccccccccc}	 
\noalign{\smallskip} 
\hline 
                              &    \multicolumn{2}{c}{[O\,II]$\lambda \lambda$3726,29}       &        &  \multicolumn{2}{c}{[O\,III]$\lambda$4363 }    & 	 &     \multicolumn{2}{c}{[O\,III]$\lambda$5007 }		  &	       &      \multicolumn{2}{c}{ [N\,II]$\lambda$6584} 			&			     &      \multicolumn{2}{c}{ [S\,II]$\lambda \lambda$6716+31} & Reference \\
\cline{2-3}
\cline{5-6}
\cline{8-9}
\cline{11-12}
\cline{14-15}
Object                        &          Obs.                &       Mod.                    &        & 	 Obs.	    & 	    Mod. 	       &    	          &	    Obs.		       &	 Mod.		  &		&	Obs.			       &	Mod.		        &			     &       Obs.		    &	  Mod.	        &       \\
\hline
IZw\,92                       &           2.63               &       2.65                    &        & 	0.32 	    &	  0.08   	       & 	 	  &	    10.12		       &	  9.19  	  &		&	  0.97  		       &	  1.01  		&			     &        0.77		    &	   0.74 	&  1	\\		   
NGC\,3393                     &           2.41               &       2.61                    &        & 	0.14 	    &	  0.09   	       & 		  &	    16.42		       &	 13.15  	  &		&	  4.50  		       &	  4.42  		&			     &        1.53		    &	   1.37 	&  2	\\
Mrk\,3                        &           3.52               &       3.72                    &        & 	0.24 	    &	  0.12   	       & 		  &	    12.67		       &	 10.64  	  &		&	  3.18  		       &	  3.25  		&			     &        1.55		    &	   1.32 	&  3	\\  
Mrk\,573                      &           2.92               &       3.07                    &        & 	0.18 	    &	  0.08   	       & 		  &	    12.12		       &	 10.02  	  &		&	  2.47  		       &	  2.52  		&			     &        1.55		    &	   1.33 	&  3	\\
Mrk\,78                       &           4.96               &       4.19                    &        & 	0.14 	    &	  0.14   	       & 		  &	    11.94		       &	 10.11  	  &		&	  2.32  		       &	  2.75  		&			     &        1.29		    &	   1.26 	&  3	\\
Mrk\,34                       &           3.43               &       3.60                    &        & 	0.15 	    &	  0.11   	       & 		  &	    11.46		       &	 10.11  	  &		&	  2.18  		       &	  2.26  		&			     &        1.62		    &	   1.43 	&  3	\\
Mrk\,1                        &           2.78               &       2.89                    &        & 	0.21 	    &	  0.11   	       & 		  &	    10.95		       &	  9.86  	  &		&	  2.21  		       &	  2.31  		&			     &        1.01		    &	   0.87 	&  3	\\
3c433                         &           6.17               &       5.73                    &        &         0.31        &     0.13                 &                  &          9.44                      &          9.24            &             &         5.13                         &          5.02                  &                            &        2.71                  &      2.61         &  3	\\
Mrk\,270                      &           5.64               &       5.56                    &        & 	0.28 	    &	  0.10   	       & 		  &	     8.71		       &	  8.18  	  &		&	  2.93  		       &	  2.74  		&			     &        2.60		    &	   2.39 	&  3	\\
3c452                         &           4.81               &       5.02                    &        & 	0.18 	    &	  0.08   	       & 		  &	     6.85		       &	  6.52  	  &		&	  3.58  		       &	  3.60  		&			     &        1.87		    &	   1.84 	&  3	\\   
Mrk\,198                      &           2.51               &       2.60                    &        & 	0.12 	    &	  0.03   	       & 		  &	     5.56		       &	  5.49  	  &		&	  2.26  		       &	  2.14  		&			     &        1.57		    &	   1.57 	&  3	\\   
Mrk\,6                        &           2.45               &       2.70                    &        & 	0.28 	    &	  0.08   	       & 		  &	    10.13		       &	  9.12  	  &		&	  1.79  		       &	  1.68  		&			     &        1.25		    &	   1.17 	&  4    \\     
ESO\,138\,G1                  &           2.35               &       2.24                    &        &         0.34        &     0.15                 &                  &         8.71                       &         8.19             &             &         0.68                         &          0.70                  &                            &        0.95                  &      0.93         &  5	\\
NGC\,3081                     &           2.16               &       2.18                    &        & 	0.23 	    &	  0.10   	       & 		   &	     12.62		       &	 10.92  	  &		&	  2.33  		       &	  2.32  		&			     &        1.22		    &	   1.77 	&  6	\\ 
NGC\,3281                     &           2.33               &       2.32                    &        &         0.24        &     0.05                 &                  &          7.59                      &         7.85             &             &         2.54                         &          2.60                  &                            &        1.13                  &      1.12         &  6	\\
NGC\,4388                     &           2.68               &       2.69                    &        & 	0.15 	    &	  0.12   	       & 		   &	     10.63		       &	 10.52  	  &		&	  1.44  		       &	  1.46  		&			     &        1.28		    &	   1.24 	&  6	\\
NGC\,5135                     &           2.01               &       1.94                    &        & 	0.10 	    &	  0.01   	       & 		   &	      4.47		       &	  4.57  	  &		&	  2.35  		       &	  2.22  		&			     &        0.72		    &	   0.77 	&  6	\\
NGC\,5728                     &           3.41               &       3.21                    &        &         0.34        &     0.11                 &                  &           10.98                    &          10.01           &             &         3.71                         &          3.74                  &                            &        0.82                  &      0.76         &  6	\\
IC\,5063                      &           5.06               &       4.63                    &        & 	0.28 	    &	  0.15   	       & 		   &	     10.31		       &	 10.25  	  &		&	  2.67  		       &	  2.62  		&			     &        1.29		    &	   1.36 	&  6	\\
IC\,5135                      &           4.05               &       3.34                    &        &         0.19        &     0.09                 &                  &           6.88                     &         7.19             &             &         3.30                         &          3.12                  &                            &        0.95                  &      1.09         &  6	\\
Mrk\,744                      &           2.38               &       2.51                    &        &         0.33        &     0.06                 &                  &           8.84                     &         8.60             &             &         3.62                         &          3.20                  &                            &        5.66                  &      5.41         &  7	\\
NGC\,5506                     &           2.84               &       2.76                    &        & 	0.14 	    &	  0.05   	       & 		   &	      7.69		       &	  7.02  	  &		&	  2.53  		       &	  2.37  		&			     &        1.91		    &	   1.70 	&  8    \\									     
Akn\,347                      &           2.98               &       3.03                    &        & 	0.42 	    &	  0.16   	       & 		   &	     15.01		       &	 15.18  	  &		&	  3.23  		       &	  3.24  		&			     &        1.50		    &	   1.43 	&  9	\\
UM\,16                        &           2.90               &       2.92                    &        & 	0.22 	    &	  0.18   	       & 		   &	     14.00		       &	 13.34  	  &		&	  1.70  		       &	  1.81  		&			     &        0.90		    &	   0.85 	&  9	\\
Mrk\,533                      &           1.59               &       1.61                    &        & 	0.13 	    &	  0.07   	       & 		   &	     12.23		       &	 11.94  	  &		&	  2.72  		       &	  2.83  		&			     &        0.84		    &	   0.79 	&  9    \\
Mrk\,612                      &           1.88               &       1.82                    &        & 	0.17 	    &	  0.06   	       & 		   &	      9.37		       &	  9.76  	  &		&	  3.60  		       &	  3.43  		&			     &        1.29		    &	   1.44 	&  9    \\
\hline
\end{tabular}
\begin{minipage}[l]{16.5cm}
 References: (1) \citet{kraemer94}, (2) \citet{contini12}, 
(3) \citet{koski78}, (4) \citet{cohen83}, (5) \citet{alloin92},  (6) \citet{phillips83}, (7) \citet{goodrich83},
(8) \citet{shuder80}, and (9) \citet{shuder81}.
\end{minipage}
\end{table*}

\begin{table*}\small
\caption{Physical parameters: $t_{3}$, $\rm O^{+}/H^{+}$, $\rm O^{2+}/H^{+}$, ICF(O), and O/H estimated
for the AGN sample (see Sect.~\ref{obs}) by using the $T_{\rm e}$-method 
(see Section~\ref{temeth}) referred as Meas.\ and the ones predicted by the  
detailed photoionization models built by \citet{dors17} (see Sect.~\ref{model}),  referred as Mod.\
 Electron density values were calculated using the observational [\ion{S}{ii}]$\lambda 6716/\lambda 6731$  line ratio (see Sect.~\ref{obs}).}
\label{tab2}
\begin{tabular}{lccccccccccccccc}	 
\noalign{\smallskip} 
\hline 
                              &          \multicolumn{2}{c}{$t_{3}$}        &               & \multicolumn{2}{c}{$12+\log(\rm O^{+}/H^{+})$}& 	         &     \multicolumn{2}{c}{$12+\log(\rm O^{2+}/H^{+})$}	 &	         &     \multicolumn{2}{c}{ICF(O)}    &      &  \multicolumn{2}{c}{$12+\log(\rm O/H)$}    &   $N_{\rm e}$ ($\rm cm^{-3}$)	\\
\cline{2-3}
\cline{5-6}
\cline{8-9}
\cline{11-12}
\cline{14-15}
Object                       &         Meas.                 &      Mod.     &               & 	 Meas.    &       Mod. 	      &    	        &	  Meas.	               &     Mod.         		 &	       &	Meas.	        &    Mod.  & 	  &  Meas.    & 	   Mod. 	          &			       \\
\hline																								        	 					        	 			   
IZw\,92         	     &           1.9509 	     &     1.0889    &  	     &   7.37	 &	 7.86	     &  		&	   7.75 	       &     8.35	  		 &	       &       1.22	   	&   1.25   &	   & 7.99      &	   8.57 		   &	   974  	      \\  
NGC\,3393       	     &           1.0939  	     &     0.9885    &  	     &   7.96	 &	 8.27	     &  		&	   8.60 	       &     8.58	  		 &	       &       1.00$^{a}$  	&   1.51   &	   & 8.68      &	   8.94 		   &	  2083  		   \\  
Mrk\,3          	     &           1.4888  	     &     1.1830    &  	     &   7.72	 &	 7.98	     &  		&	   8.11 	       &     8.24	  		 &	       &       1.23	   	&   1.65   &	   & 8.35      &	   8.65 		   &	  1059  		   \\  
Mrk\,573        	     &           1.3371  	     &     1.0399    &  	     &   7.73	 &	 8.09	     &  		&	   8.21 	       &     8.41	  		 &	       &       1.39	   	&   1.48   &	   & 8.48      &	   8.75 		   &	   876  		   \\  
Mrk\,78         	     &           1.2191  	     &     1.2735    &  	     &   8.02	 &	 7.88	     &  		&	   8.32 	       &     8.11	  		 &	       &       1.39	   	&   1.79   &	   & 8.64      &	   8.56 		   &	   396  		   \\  
Mrk\,34         	     &           1.2709  	     &     1.1402    &  	     &   7.83	 &	 7.99	     &  		&	   8.25 	       &     8.28	  		 &	       &       1.25	   	&   1.57   &	   & 8.49      &	   8.65 		   &	   596  		   \\  
Mrk\,1          	     &           1.4974  	     &     1.1641    &  	     &   7.60	 &	 7.87	     &  		&	   8.04 	       &     8.25	  		 &	       &       1.32	   	&   1.48   &	   & 8.30      &	   8.58 		   &	   863  		    \\ 
3c433                        &           1.9951  	     &     1.2764    &  	     &   7.64	 &	 7.75	     &  		&	   7.70 	       &     8.05	  		 &	       &       1.00$^{a}$  	&   2.14   &	   & 7.97      &	   8.64 		   &	    10  		    \\ 
Mrk\,270        	     &           1.9700  	     &     1.2105    &  	     &   7.71	 &	 8.05	     &  		&	   7.68 	       &     8.04	  		 &	       &       1.12	   	&   1.74   &	   & 8.05      &	   8.59 		   &	  1227  		    \\ 
3c452           	     &           1.7569  	     &     1.2387    &  	     &   7.62	 &	 7.95	     &  		&	   7.68 	       &     7.94	  		 &	       &       1.03	   	&   1.72   &	   & 7.97      &	   8.48 		   &	    10  		    \\ 
Mrk\,198        	     &           1.5857  	     &     0.9183    &  	     &   7.44	 &	 8.17	     &  		&	   7.69 	       &     8.35	  		 &	       &       1.07	   	&   1.30   &	   & 7.91      &	   8.69 		   &	   118  		    \\ 
Mrk\,6          	     &           1.8065  	     &     1.0665    &  	     &   7.38	 &	 7.96	     &  		&	   7.82 	       &     8.35	  		 &	       &       1.26	   	&   1.40   &	   & 8.06      &	   8.65 		   &	   794  		    \\ 
ESO\,138\,G1                 &           2.2192  	     &     1.4387    &  	     &   7.22	 &	 7.48	     &  		&	   7.58 	       &     7.90	  		 &	       &       1.29	   	&   1.71   &	   & 7.85      &	   8.77 		   &	   794  		    \\ 
NGC\,3081       	     &           1.4623  	     &     1.0715    &  	     &   7.52	 &	 7.90	     &  		&	   8.13 	       &     8.41	  		 &	       &       1.30	   	&   1.55   &	   & 8.34      &	   8.72 		   &	   932  		    \\ 
NGC\,3281                    &		 1.9509		     &	   0.9354    &		     &	 7.33	 &	 7.93	     &  		&	   7.62 	       &     8.48		  	 &	       &       1.00$^{a}$    	&   1.29   &	   & 7.79      &	   8.77 		   &	 1126			   \\ 
NGC\,4388       	     &           1.3094  	     &     1.1694    &  	     &   7.67	 &	 7.81	     &  		&	   8.18 	       &     8.27	  		 &	       &       1.13	   	&   1.57   &	   & 8.35      &	   8.60 		   &	  364			   \\ 
NGC\,5135       	     &           1.6145  	     &     0.7816    &  	     &   7.37	 &	 8.43	     &  		&	   7.57 	       &     8.55	  		 &	       &       1.11	   	&   1.20   &	   & 7.83      &	   8.88 		   &	  551			   \\ 
NGC\,5728                    &           1.9272  	     &     1.1508    &  	     &   7.46	 &	 7.66	     &  		&	   7.80 	       &     8.21	  		 &	       &       1.11	   	&   1.53   &	   & 8.01      &	   8.62 		   &	  650			   \\ 
IC\,5063        	     &           1.7890  	     &     1.3182    &  	     &   7.67	 &	 7.87	     &  		&	   7.84 	       &     8.05	  		 &	       &       1.09	   	&   1.96   &	   & 8.10      &	   8.56 		   &	  365			   \\ 
IC\,5135                     &           1.8056  	     &     1.2246    &  	     &   7.57	 &	 8.23	     &  		&	   7.65 	       &     8.26	  		 &	       &       1.22	   	&   1.48   &	   & 8.00      &	   8.88 		   &	  514			   \\ 
Mrk\,744                     &           2.1575  	     &     0.9268    &  	     &   7.23	 &	 8.30	     &  		&	   7.61 	       &     8.58	  		 &	       &       1.00$^{a}$  	&   1.69   &	   & 7.76      &	   8.97 		   &	  725			   \\ 
NGC\,5506       	     &           1.4616  	     &     1.0092    &  	     &   7.64	 &	 8.10	     &  		&	   7.91 	       &     8.32	  		 &	       &       1.18	   	&   1.30   &	   & 8.17      &	   8.64 		   &	  932			    \\
Akn\,347        	     &           1.8189  	     &     1.1324    &  	     &   7.44	 &	 8.03	     &  		&	   7.98 	       &     8.44	  		 &	       &       1.29	   	&   1.94   &	   & 8.21      &	   8.87 		   &	  564			    \\
UM\,16          	     &           1.3693  	     &     1.2416    &  	     &   7.69	 &	 7.79	     &  		&	   8.25 	       &     8.27	  		 &	       &       1.23	   	&   1.78   &	   & 8.44      &	   8.65 		   &	  747			    \\
Mrk\,533        	     &           1.1769  	     &     0.9183    &  	     &   7.62	 &	 8.12	     &  		&	   8.37 	       &     8.68	  		 &	       &       1.34	   	&   1.52   &	   & 8.57      &	   8.97 		   &	 1131			   \\ 
Mrk\,612        	     &           1.4593 	     &     0.8974    &  	     &   7.38	 &	 8.19	     &  		&	   8.00 	       &     8.61	  		 &	       &       1.15	   	&   1.66   &	   & 8.16      &	   8.97 		   &	   51			  \\  
\hline																															   
\end{tabular}																														  
\begin{minipage}[l]{18cm}
$^{a}$ ICF(O) assumed to be equal to 1.0 as explained in the text.
\end{minipage}
\end{table*}

\begin{figure}
\centering
\includegraphics[angle=-90,width=1\columnwidth]{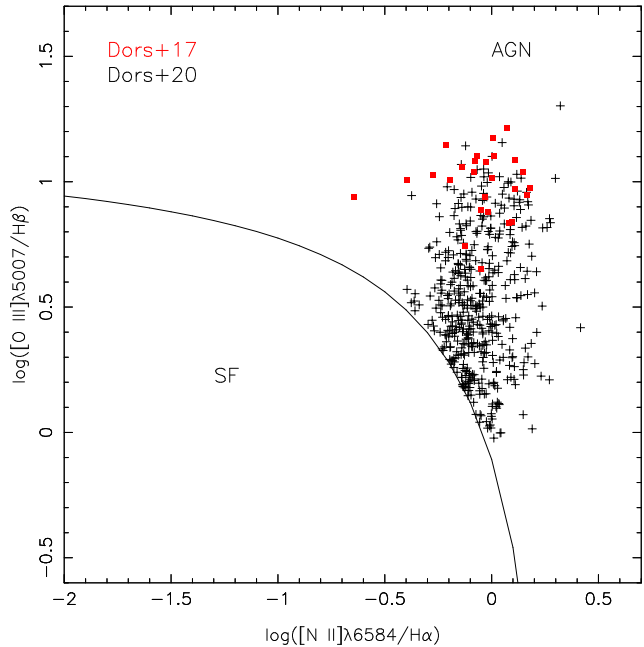}
\caption{[\ion{O}{iii}]$\lambda$5007/H$\beta$ versus [\ion{N}{ii}]$\lambda$6584/H$\alpha$ diagnostic diagram for the objects 
in our samples (see Sect.~\ref{meth}). Each sample is represented with a different colour as indicated.
Solid line, taken from \citet{kewley01}, separates SF-like objects from AGN-like objects.}
\label{discf1}
\end{figure}

\begin{figure*}
\centering
\includegraphics[angle=-90,width=1\columnwidth]{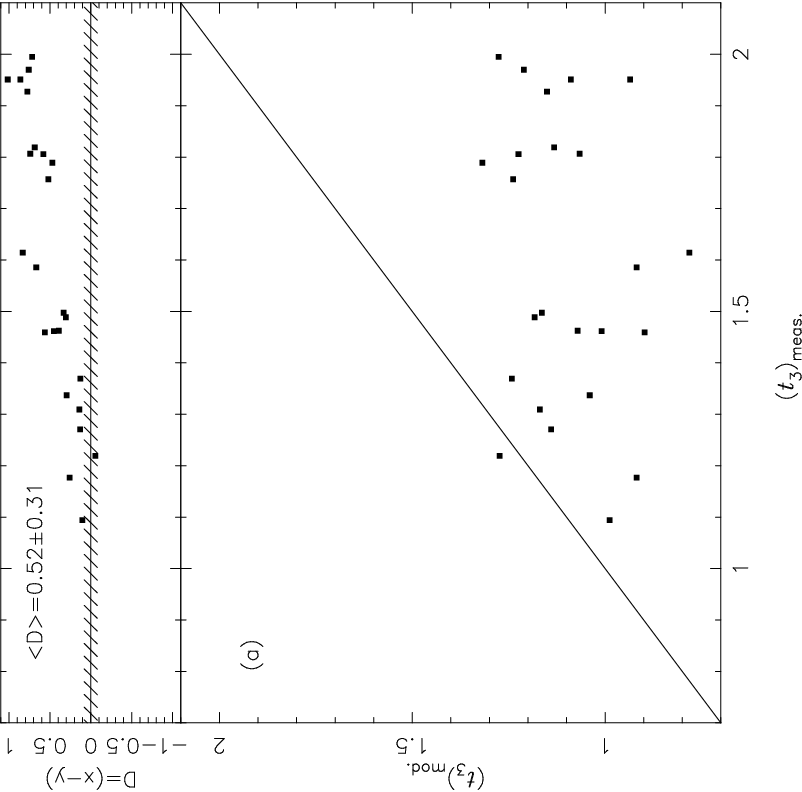}
\includegraphics[angle=-90,width=1\columnwidth]{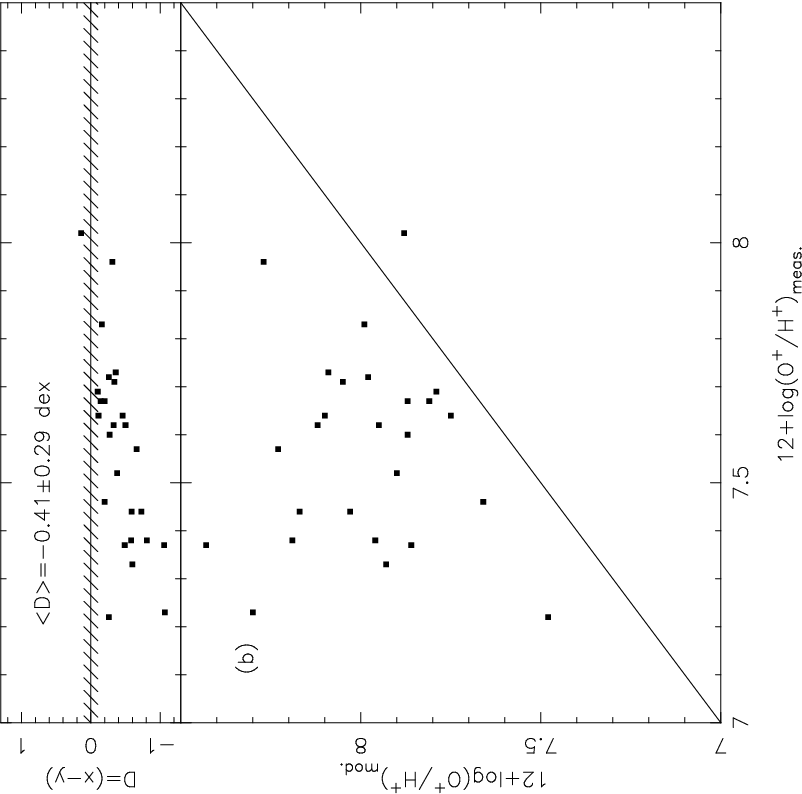}\\[5pt]
\includegraphics[angle=-90,width=1\columnwidth]{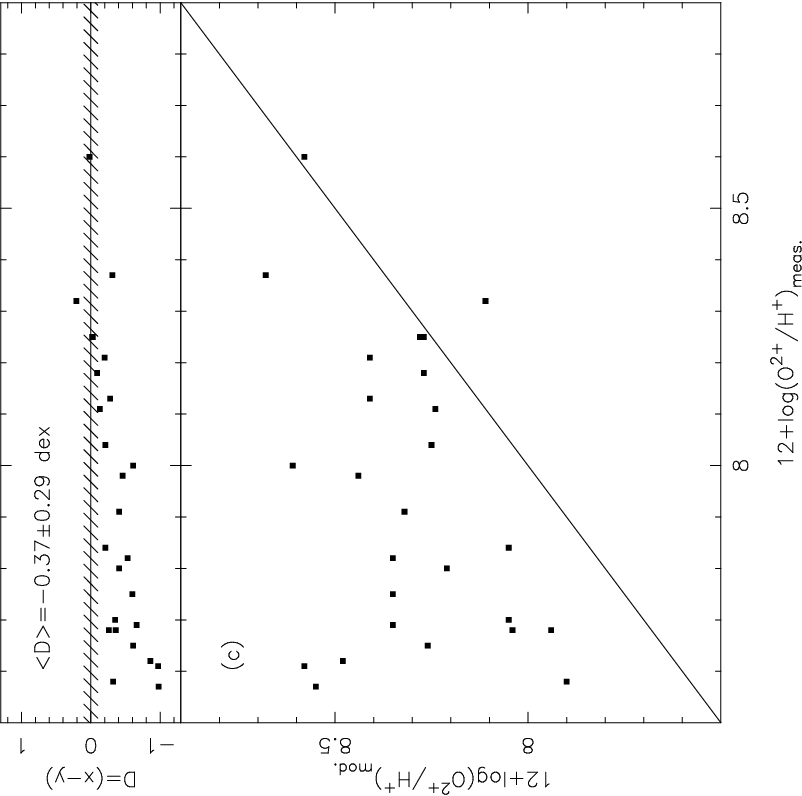}
\includegraphics[angle=-90,width=1\columnwidth]{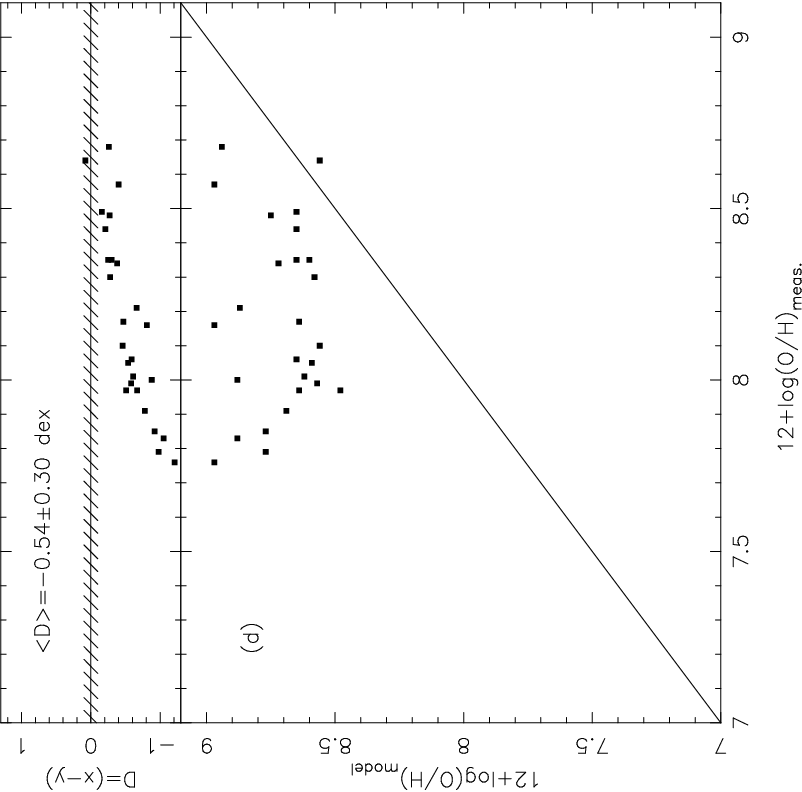}
\caption{Physical parameter comparison for the objects in the \citet{dors17} sample. Panel (a): Points in the bottom sub-panel represent  $t_{3}$ values
predicted by detailed photoionization models [$(t_{3})_{\rm mod.}$; 
see Sect.~\ref{model}] versus those calculated through the $T_{\rm e}$-method [$(t_{3})_{\rm meas.}$; see Sect.~\ref{temeth}]. 
Solid line represents the equality between both estimations. In top
sub-panel the difference between these estimations (D=x-y) is shown.  
Panel (b): As in panel (a) but for $12+\log(\rm O^{+}/H^{+})$. 
Panel (c): As in panel (a) but for $12+\log(\rm O^{2+}/H^{+})$.
Panel(d): As in panel (a) but for the total oxygen abundance $12+\log(\rm O/H)$.
In each panel the averaged diference $\rm <D>$ is shown. 
The hatched area in panel (a) represents the $\pm 0.08$ uncertainty 
in the $t_{3}$ values derived by \citet{kennicutt03} using the $T_{\rm e}$-method, 
and in panels (b), (c) and (d) it represents the uncertainty of 0.1 dex in abundances 
derived through $T_{\rm e}$-method (see \citealt{kennicutt03, hagele08}). }
\label{f1}
\end{figure*}

\section{Methodology}
\label{meth}

With the aim to study the different factors that could contribute to the discrepancy between O/H as derived 
from the $T_{\rm e}$-method and from models in
Seyfert 2 nuclei, we used the two observational samples 
taken from the literature and the detailed photoionization models considered
by \citet{dors17},  described in the following subsections.

% _____________________________________________

\subsection{Observational data}
\label{obs}

\subsubsection{\citet{dors17} sample}
\label{sdors17}

Optical narrow  emission-line intensities of AGNs classified as Seyfert 2 and 1.9 compiled by
\citet{dors17} were used as observational data. The data include de-reddened flux measurements of  [\ion{O}{ii}]$\lambda$3726+$\lambda$3729 
(referred as [\ion{O}{ii}]$\lambda$3727), 
[\ion{O}{iii}]$\lambda$4363,  H$\beta$, [\ion{O}{iii}]$\lambda$5007,  H$\alpha$, [\ion{N}{ii}]$\lambda$6584, and
[\ion{S}{ii}]$\lambda\lambda$6717,31  of 47 local AGNs (redshift $z<0.1$).
It is possible to apply the $T_{\rm e}$-method to 26 of these objects which constitute our final observational sample,
 hence the other objects present intensities of the [\ion{O}{iii}]($\lambda 4959+\lambda 5007)$)/$\lambda 4363$
line ratio out of the range of  permited values (see below) in the calculation of the electron temperature.
The emission-lines have Full Width Half Maximum  (FWHM)  lower than 1000 $\rm km \: s^{-1}$, what 
indicates that they are produced in the NLRs where the gas shock
has little influence on the heating and ionization.  
In addition,  electron densities ($N_{\rm e}$) derived from [\ion{S}{ii}] ratio in
NLRs are found in leading to the low density regime,  with   $ N_{\rm e}\: \la \: 1000 \: \rm cm^{-3}$ 
for most objects (see \citealt{vaona12, zhang13, dors14}),
where the colisional de-excitation has a negligible effect on emission-line formation. 
In Table~\ref{tab1}, the reddening corrected emission-line intensities (relative to H$\beta$=1.0) 
are listed.

Although the compiled data constitute a heterogeneous sample, e.g.\
they were obtained using different observational techniques and measurement apertures,
the effects of using such data do not yield any bias on the abundance estimations
(see a complete discussion about these points in \citealt{dors13, dors20}).    

% ______________________________________________________

\subsubsection{\citet{dors20} sample}
\label{sdors20}

 To analyse the effect of the new formalism of the $T_{\rm e}$-method (see below) on the O/H abundances of NLRs,
we also taken into account a sample of  463 confirmed  Seyfert~2 nuclei compiled by \citet{dors20}.
This large sample was obtained performing a cross-correlation between the  Sloan Digital Sky Survey (SDSS, \citealt{york00}) and the 
NASA/IPAC Extragalactic Database (NED) to selected
optical ($3000 \: < \: \lambda({\mathrm{\AA}}) \: < \: 7000$) emission line intensities of   
Seyfert 2  nuclei with redshift $z \: \la \: 0.4$. The reader is referred to  \citet{dors20} for a complete description 
of the observational data. 

In Figure~\ref{discf1}, we show the location of the objects in the \citet{dors17} and  \citet{dors20} samples in the
 [\ion{O}{iii}]$\lambda$5007/H$\beta$ versus [\ion{N}{ii}]$\lambda$6584/H$\alpha$
diagnostic diagram, often used to distinguish   star forming galaxies  from AGNs.
In this figure we also included the line proposed by
\citet{kewley01} to separate the two objects classes.
It can be seen that all the objects in our samples are in the AGN-like region of the diagram,
therefore, these objects are appropriated for the analysis in this work hence their 
spectra are dominated by the NLR emission.

\subsection{$T_{\rm e}$-method: H{\scriptsize II} region formalism}
\label{temeth}

 For the \citet{dors17} sample, we compute the ionic abundance ratios
$\rm O^{+}/H^{+}$ and $\rm O^{2+}/H^{+}$ as well as 
total oxygen abundance (O/H) adopting   the equations given by 
\citet{enrique03}, \citet{enrique09} and \citet{hagele08}. These equations
are the same ones considered by \citet{dors15}.

The electron temperature for the high ionization zone (refered as $t_{3}$) was calculated
from the observed line-intensity ratio 
$R_{\rm O3}$=(1.33$\times$$I$[\ion{O}{iii}]$\lambda 5007)$/$I$[\ion{O}{iii}]$\lambda4363$
using the expression
\begin{equation}
 \label{eqt3}
 t_{3}=0.8254-0.0002415 R_{\rm O3}+\frac{47.77}{R_{\rm O3}},
 \end{equation}
\noindent with $t$ in units of $10^{4}$ K. This relation is valid in the
range of $700 \: \ga \: R_{\rm O3}  \: \ga \: 30$ which corresponds to 
$0.7 \: \la \: t_{3} \: \la \: 2.3$ \citep{hagele08}. 
The 26 objects selected for the final sample are 
those with the estimated $t_{3}$ in the equation validity range.

The electron temperature for the low ionization zone (refered as $t_{2}$)
was estimated using the expresssion:
\begin{equation}
\label{eqt2}
t_{2}^{-1}\,=\,0.693\,t_{3}^{-1}+0.281.
\end{equation}
This  relation was derived using $t_{2}$ and $t_{3 }$ values predicted by 
photoionization models simulating \ion{H}{ii} regions.

The electron density ($N_{\rm e}$), for each object of the sample, was calculated  from 
the [\ion{S}{ii}]$\lambda 6716/\lambda 6731$ emission-line intensity ratio by 
using the {\sc IRAF/temden} task, with  the $t_{2}$ values calculated from Eq.~\ref{eqt2}.

The $\rm O^{2+}$ and $\rm O^{+}$ abundances were computed using the
following relations: 
\begin{eqnarray}
\label{eqt4}
 12+\log(\frac{{\rm O^{2+}}}{{\rm H^{+}}}) \!\!\!&=&\!\!\! \log \big( \frac{1.33 \times I(5007)}{I{\rm (H\beta)}}\big)+6.144  \nonumber\\
                                          &&\!\!\!+\frac{1.251}{t_{3}}-0.55\log t_{3} 
\end{eqnarray}
 and
\begin{eqnarray}
\label{eqt5}
 12+\log(\frac{{\rm O^{+}}}{{\rm H^{+}}})  \!\!\!&=&\!\!\! \log  \big( \frac{I(3727)}{I{\rm (H\beta)}}\big)+5.992 \nonumber\\
                                           &&\!\!\!+\frac{1.583}{t_{2}}-0.681\log t_{2} +\log(1+2.3 n_{\rm e}),
\end{eqnarray}
where $n_{\rm e}$ is the electron density $N_{\rm e}$ in  units of 10\,000 $\rm cm^{-3}$.

Finally, the total oxygen abundance (O/H) was computed assuming
\begin{equation}
\label{eqt6}
{\rm
\frac{O}{H}=ICF(O)\: \times \: \left[\frac{O^{2+}}{H^{+}}+\frac{O^{+}}{H^{+}}\right],} 
\end{equation}
where  ICF(O) is the Ionization Correction Factor for oxygen that take into account the contribution of the unobservable oxygen ions.
 We consider the ICF(O) expression proposed for Planetary Nebula (PN)  by \citet{torrespeimber77}
\begin{equation}
\label{icfox}
\rm
ICF(O)=\frac{N(He^{+}+He^{2+})}{N(He^{+})},
\end{equation}
where N represents the abundance. This ICF expression is based on the similarity between
the $\rm He^{+}$ and $\rm O^{2+}$ ionization potential (about 54 eV) and it can be applied for any object class.
To calculate the ionic helium abundance for each object,  we consider the expressions by \citet{izotov94}:
\begin{equation}
\label{abundhe}
 {\rm \frac{N(He^{+})}{N(H^{+})}}=0.738 \: t^{0.23} \: \frac{I(\lambda 5876)}{I(\rm H\beta)}
\end{equation}
and
\begin{equation}
 {\rm \frac{N(He^{2+})}{N(H^{+})}}=0.084 \: t^{0.14} \: \frac{I(\lambda 4686)}{I(\rm H\beta)},
\end{equation}
where  $t=t_{3}$  was assumed. It was not possible to calculate the ICF(O) for four objects of
the sample: NGC 3393, 3c433, NGC 3281 and  Mrk 744, because the \ion{He}{ii}$\lambda 4686$  
emission-line is not listed in the original works from which the data were compiled. For these objects ICF(O)=1.0 was assumed.
Typical errors in the emission-line intensity measurements for the objects 
in our sample are of the order of about 10-20 per cent (e.g.\ \citealt{kraemer94}), which
translate into O/H abundance uncertainties of about 0.1 dex (e.g. \citealt{kennicutt03, hagele08}).

\subsection{Photoionization models}
\label{model}
  
Detailed photoionization models  built by \citet{dors17}   using the {\sc Cloudy} code  version 13.04  \citep{ferland13}
were used   to calculate
the ionic and total oxygen abundances as well as the ICF(O) of the objects listed in Table~\ref{tab1}.
The input parameters of the models were: metallicity, 
abundances of  the N and S elements, power law index ($\alpha$) of the Spectral Energy Distribution (SED), 
electron density ($N_{\rm e}$),   number of ionizing photons [$Q(\rm H)$] and 
inner radius ($R_{\rm in}$), defined as being  the distance from the ionizing source to the illuminated 
gas region. These nebular parameters were varied during the fitting procedure accordingly to the {\sc phymir} 
optimize method \citep{vanhoof97}. As usual, the oxygen abundance O/H was scaled linearly with the
metallicity, while the N/H and S/H abundances were considered free parameters,  i.e no fixed
relation between the abundances of these elements and  O/H was assumed during the fitting.
The outermost radius ($R_{\rm out}$) was defined as the one where the temperature reaches 4\,000 K, 
a default procedure in the {\sc Cloudy} code. We carried out several simulations considering larger values 
of $R_{\rm out}$ (e.g. stopping the calculations in the region with  $T_{\rm e}$=1\,000 K) in order to consider
the emission from the neutral gas, necessary to reproduce molecular emission of AGNs (see \citealt{dors12, riffel13}).
We found that the intensity of the predicted optical lines are practically
the same as  those considering the default value of $R_{\rm in}$.

Basically, from  a series of models, \citet{dors17} selected the best model to describe 
the observed emission-line flux ratios of a specific AGN. This model produces 
the smallest value of  $\chi_{T}$, with 
\begin{equation}
\chi_{T}=\sum \chi_{i}=\sum (I_{\rm obs.}^{i}-I_{\rm pred.}^{i})^{2}/I_{\rm obs.}^{i},
\end{equation}
where $I_{\rm obs.}^{i}$ and $I_{\rm pred.}^{i}$ are the observational and predicted
intensities of the line ratio $i$, respectively. 
The   difference between $I_{\rm obs.}^{i}$ and $I_{\rm pred.}^{i}$ was required to be 
lower than $20 \%$, which is a typical observational uncertainty for
emission lines (e.g. \citealt{kraemer94}).  \citet{dors17}
performed several simulations in order
to reproduce the observational intensity of [\ion{O}{iii}]$\lambda$4363/H$\beta$, i.e. considering models
with fluctuations of metallicity  and electron density. Also, these authors adopted the same
methodology used by \citet{dors15}, which  young stellar clusters,  whose spectra were computed with the $STARBURST99$ code \citep{leitherer99},
are considered as  secondary ionization source.
However, for the few cases which  was possible reproduce this line ratio, 
other emission-lines (e.g. [\ion{O}{iii}]$\lambda$3727, [\ion{O}{iii}]$\lambda$5007)
were not reproduced by the models. Therefore, the requirement above was not applied for  [\ion{O}{iii}]$\lambda$4363.
The uncertainty in the model resulting elemental abundances  found is $\sim 0.1$ dex.
 Similar photoionization model fitting was considered by \citet{contini17a, contini17b} in order
to reproduce observational line ratio intensities of SFs, AGNs and gamma-ray burst host galaxies.
In Table~\ref{tab1}, the model-predicted emission-line intensities are compared to the
observational ones. It can be seen there is a good agreement between them, with exception
of the [\ion{O}{iii}]$\lambda$4363/H$\beta$  ratio, for which the observational value is about 2.5 
times higher than the predicted one. It is worth to be mentioning that for Mrk\,78, Mrk\,34, NGC\,4388, and
Mrk\,533 this observational ratio is reproduced by the models taking into account the observational
uncertainty of 20\%. The ICF(O) for each object were computed from the photoionization model fittings assuming the expression
\begin{equation}
\rm
ICF(O)=\frac{N(O)}{N(O^{+}+O^{2+})}.
\end{equation}

In addition, a grid of photoionization models was built to obtain the electron temperature
predictions for regions along the AGN radius containing different ions, i.e. in order to derive
a new $t_{2}$-$t_{3}$ relation for AGNs (see below).   
This grid is similar to the one considered
by \citet{sarita20} and it covers a wide range of physical parameters.
The photoionization models
assume as SED  a multicomponent continuum with the usual shape and parameters values 
typical for AGNs. One of these SED parameters is the slope of the power law ($\alpha_{ox}$) 
proposed to model the continuum between 2500\AA\ (in the UV) and 2 keV (in X-rays). As changes
 in this slope imply changes in the hardness of the source radiation, we assume three values
  for this parameter:  $-$0.8, $-1.1$ and $-1.4$.  
The logarithm of the ionization parameter ($U$) was considered
in the range  $ -3.5 \: \lid \: \log U \: \lid \: -0.5$, with a step of 0.5 dex.  The metallicity was assumed
to take the following values ($Z/Z_{\odot}$)= 0.2, 0.5,  0.75, 1.0, and 2.0.
 Metallicities in this range has been found   for local AGNs and out to $z\sim7$
(e.g. \citealt{nagao06a, feltre16, matsuoka18, thomas19, mignoli19, enrique19, dors20, sarita20}).
The nitrogen and oxygen abundance relation: 
$\rm log(N/O)=1.29\times [12+log(O/H)] - 11.84$, obtained using the estimations by \citet{dors17} for \ion{H}{ii} regions and AGNs, 
was assumed in the models. Four values of electron density, $N_{\rm e}$=100, 500, 1500 and 3000 $\rm cm^{-3}$, were considered
in the models.   The predicted $t_{2}$, $t_{3}$ and  ionic abundance values  are  those weighted over 
nebulae volume times electron density of the models.

% ______________________________________________

\section{Results}
\label{res}
 
In Table~\ref{tab2},   the estimations  through the $T_{\rm e}$-method and from  
detailed photoionization model predictions of $t_{3}$, $12+\log(\rm O^{+}/H^{+})$, $12+\log(\rm O^{2+}/H^{+})$, ICF(O),  the total
oxygen abundance $12+\log(\rm O/H)$, and the $N_{\rm e}$ (calculated via the observational
[\ion{S}{ii}]$\lambda$6716/$\lambda$6731 emission-line ratio) for each object in our sample are listed.

%---------------------------------------------
In Fig.\ \ref{f1},   the results  obtained from the $T_{\rm e}$-method versus those derived  from 
photoionization models for $t_{3}$ (panel a),  $12+\log(\rm O^{+}/H^{+})$ (panel b), $12+\log(\rm O^{2+}/H^{+})$ (panel c) 
and for the total oxygen abundance 12+log(O/H) (panel d), are shown. 
In panel (a) we can see that the 
$t_{3}$ difference increases (almost systematically from $\sim 0$ to 
$\sim 1.1$, i.e. up to $\sim 11000$ K) when the values derived by the $T_{\rm e}$-method increase. 
This difference is higher 
 than the electron temperature uncertainties of $\sim 800$ K estimated for star-forming 
 regions \cite[see e.g.][]{kennicutt03}.  In panel (b) we notice that the $\rm O^{+}/H^{+}$
 difference increases for low ionic abundances with the average difference of about 0.4 dex and rising up to $\sim 1$ dex. 
 The  $\rm O^{2+}/H^{+}$ results, shown in panel (c),
 have a similar behaviour of $\rm O^{+}/H^{+}$, with an average difference  of about 0.4 dex. 
The total oxygen abundances, O/H (panel d), derived by using the $T_{\rm e}$-method 
are (almost systematically) lower than those predicted by the models. The difference, D, between both
 estimations increases as the metallicity (traced by the O/H abundance) decreases, with 
an average difference of  $\sim 0.5$ dex. This total oxygen abundance discrepancy is somewhat lower ($\sim 0.1$ dex) than the ones
found by \citet{dors15, dors20}, 
who compared the O/H values derived through the $T_{\rm e}$-method with the values obtained 
via calibrations proposed by  \citet{thaisa98} and  \citet{castro17}.
This sligth difference between the O/H estimations is  mainly due to the fact that \citet{dors15, dors20} did not apply any ICF(O)
when the $T_{\rm e}$-method was considered. 
Finally, the ICF(O) values derived by using the He abundances (Eq.~\ref{icfox})
ranges from $\sim1.0$ to $\sim1.4$ and they are somewhat lower than those predicted by the photoionization models, found to be
  ranging  from $\sim1.2$ to $\sim 2.2$.

%----------------------------------

\section{Discussion}
\label{disc}

Chemical evolutionary models of spiral and elliptical galaxies predict, for the central parts of these objects, 
metallicities in the $0.5 \: \la \: (Z/Z_{\odot}) \: \la \: 2.0$ range (e.g. \citealt{molla05})
in agreement with observational estimations obtained by extrapolations of chemical abundance gradients 
(an independent metallicity estimation; e.g. \citealt{vila92, thaisa98, pilyugin04, igor19}).
However, \citet{dors15, dors20} found that oxygen abundance estimations based on narrow optical emission-lines emitted by type 
2 AGNs and derived through the $T_{\rm e}$-method (the most reliable method for \ion{H}{ii} regions)
are, in general, sub-solar, and are underestimated by about 0.6 dex when compared to those derived
from strong emission-lines methods and from the central intersect gradient method.
In particular, \citet{dors20}  used a large sample of Seyfert 2 nuclei  and  found an 
 average value of $\rm 12+log(O/H)\approx8.0$  or $(Z/Z_{\odot})\approx0.2$ when the $T_{\rm e}$-method
was applied.

The results found in the present work show that the total O/H abundances derived for our sample through the $T_{\rm e}$-method
are in the $\rm 7.8 \: \la \: 12+\log(O/H) \: \la \:8.70$ range
which, assuming the solar oxygen value  of 12+log(O/H)$_\odot$=8.69 \citep{allendeprieto01}, corresponds to a
metallicity range of  $0.10 \: \la \: (Z/Z_{\odot}) \: \la \: 1.0$, i.e.\ sub-solar metallicities.
On the other hand, detailed photoionization models predict O/H abundances in the range
of $\rm 8.5 \: \la \: 12+\log(O/H) \: \la \:9.0$ or $0.60 \: \la \: (Z/Z_{\odot}) \: \la \: 2.0$.
This O/H ($Z$) discrepancy  has been attributed to the fact that photoionization models predict lower temperature values than
those directly estimated from observational $R_{O3}$  ratios, the so-called $T_{\rm e}$-problem (e.g.\ \citealt{komossa97, zhang13}).
We also found a difference between $t_3$ values derived from  the $T_{\rm e}$-method and 
those predicted by photoionization models, which increases for high electron temperature values.

The origin of the  $T_{\rm e}$-problem in AGNs  is an open issue in
nebular astrophysics and it is not necessarily the same as in \ion{H}{ii} regions.
Moreover, an acceptable solution  has not been already proposed.  Subsequently,
a discussion about  possible sources of  the $T_{\rm e}$-problem is presented.

\subsection{Electron density}

\begin{figure}
\centering
\includegraphics[angle=-90,width=1\columnwidth]{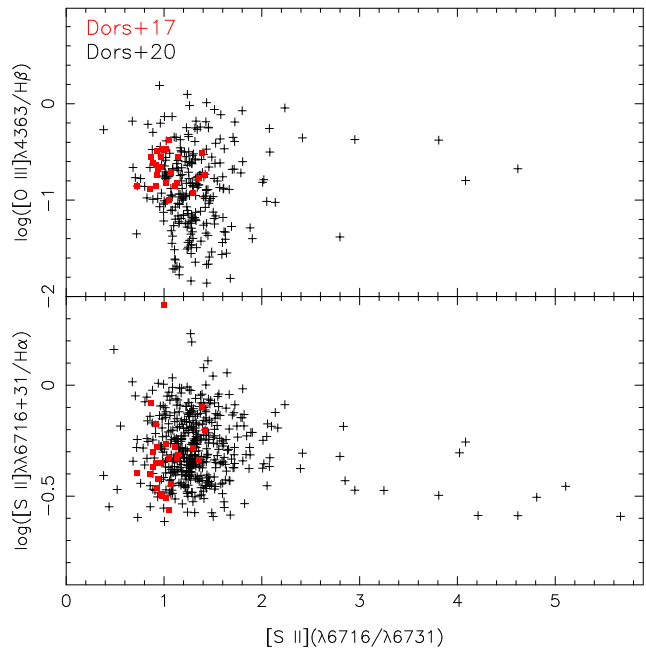}
\caption{[\ion{S}{ii}]/H$\alpha$ and [\ion{O}{iii}]$\lambda$4363/H$\beta$
versus [\ion{S}{ii}]$\lambda$6716/$\lambda$6731
for our samples of objects.  The \citet{dors17} and \citet{dors20} samples described in Sects. \ref{sdors17}
and \ref{sdors20}, respectively, are indicated by different colours.}
\label{figtr}
\end{figure}

\citet{nagao01}, using observational optical and infrared data of AGNs and photoionization model results,
presented evidences that a large fraction of [\ion{O}{iii}]$\lambda$4363 flux is  
emitted in a more dense   ($N_{\rm e}\sim 10^{5-7} \: \rm cm^{-3}$)
and obscured gas regions than those emitting  [\ion{O}{iii}]$\lambda$5007 (see also \citealt{crenshaw05, baskin05, kraemer11}), 
being NLRs composed of gas clouds with a variety of electron density (e.g. \citealt{ferguson97}).
In fact, electron density determinations based on [\ion{S}{ii}]$\lambda$6716/$\lambda$6731 
and [\ion{Ar}{iv}]$\lambda$4711/$\lambda$4740 line ratios by  \citet{congiu17} show an electron density  
stratification in the NLRs of two Seyfert 2 (IC\,5063 and NGC\,7212), with $N_{\rm e}$ ranging from
$\sim 200$ to $\sim 13\,000$ $\rm cm^{-3}$.  
\citet{izabel18} performed emission-line flux two-dimensional maps 
of five bright nearby Seyfert nuclei obtaining electron density 
variations along the central part of these objects, with $N_{\rm e}$ ranging from $\sim 100$
to $\sim 2500 \: \rm cm^{-3}$. 
\citet{revalski18a} found a density profile in the NLR of Mrk\,573,
with a peak of about 3000 $\rm cm^{-3}$ at the center and a decrease following a shallow power law with radial distance. 
\citet{kakkad18}  presented electron density maps for a sample of 13  outflowing and non-outflowing Seyfert galaxies.
These authors found  non-uniform distribution of electron densities with values varying from about 50 to 2000 $\rm cm^{-3}$.
\citet{mingozzi19} used MUSE data of nearby Seyfert 2 to map their density structure and found a broad range of densities 
from 200 to 1000 cm$^{-3}$, but mostly peaked at low densities.
However, electron density estimations based on the [\ion{S}{ii}]$\lambda$6716/$\lambda$6731
could be somewhat uncertain. For example, \citet{davies20}, using optical emission line intensities
of 11 Seyfert 2, showed that electron density derived from only the [\ion{S}{ii}] doublet
is signicantly lower (by a factor from 4 to 10) than  that derived through both auroral  and transauroral lines \citep{holt11} 
as well as by using the method based on ionization parameter determination \citep{baron19}.
The latter method is somewhat uncertain, given that the ionization parameter depends quadratically 
from the radial distance and that is only known in projection.
The highest electron density value obtained by \citet{davies20} was $\sim 67\,000 \: \rm cm^{-3}$ for the Seyfert 2 NGC\,5728
considering the  ionization parameter method. 
This value is much lower than the critical density value for the [\ion{O}{iii}]$\lambda$4363 emission-line ($N_{\rm c}=10^{7.5} \: \rm cm^{-3}$, \citealt{vaona12}).
Therefore, it is unlikely that electron density variations
are the main origin of the temperature discrepancies found here.
In any case, in order to verify if there is indication of some electron density effect on our analysis,
in Fig.~\ref{figtr}, we show the [\ion{S}{ii}]($\lambda\lambda$6716, 31)/H$\alpha$ and [\ion{O}{iii}]$\lambda$4363/H$\beta$
versus [\ion{S}{ii}]$\lambda$6716/$\lambda$6731 emission-lines ratios for our both samples,  indicating that there is no correlation.
We neither found any correlation between the [\ion{O}{iii}]$\lambda$4363/H$\beta$ ratio 
and the  $\rm C(H\beta)$ extinction coefficient (not shown).
It is worth to be mention that in Fig.~\ref{figtr} there are several objects\footnote{These objects are not
considered in our O/H estimations.} from the SDSS sample presenting unphysically large values of the sulfur emission-line ratio, 
i.e.\ values larger than 1.42 that is the theoretical value for the low density limit \citep{Osterbrock06}. Similar results 
were already found for \ion{H}{ii}-regions and \ion{H}{ii}-galaxies
using different kind of instruments \citep{kennicutt89,zaritsky94,lagos09,relanio10,lopezh13,krabbe14}.
It was suggested by \citet{lopezh13} that such high ratio values could be due to some problem in the sulphur atomic data. 
They suggested that when the sulphur ratio is above the 1.42 limit, a safe way to proceed is to assume an  electron density of 
100 cm$^{-3}$ since even before reaching this theoretical limit the density estimations are very uncertain. This procedure is also followed by \citet{krabbe14}.
In reference to the clumps of very high density that could be present in NLRs, they are still not detected, 
for instance,
in Integral Field Unit studies as the ones carried out by \citet{izabel18}\footnote{The spatial resolution 
of \citet{izabel18} observations ranges from 110 to 280 pc.} and by \citet{mingozzi19}. We therefore conclude 
 that  electron density variations do not have a significant effect on the formation of 
the emission-lines and, consequently, on the $T_{\rm e}$-method use in NLRs.

\begin{figure}
\centering
\includegraphics[angle=-90,width=1\columnwidth]{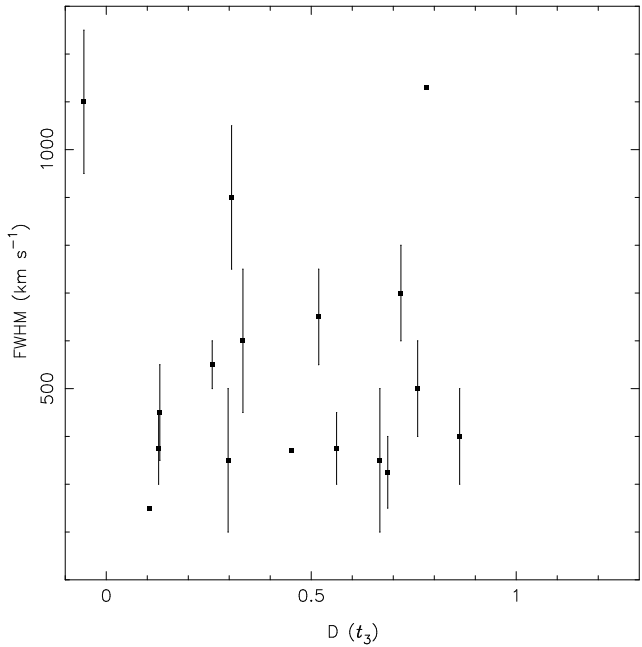}
\caption{FWHM (in $\rm km \: s^{-1}$) of permitted emission lines (H$\beta$ or H$\alpha$) of objects in the \citet{dors17} sample
versus the difference between $t_{3}$ values calculated by the $T_{\rm e}$-method and those predicted by detailed photoionization models
obtained from Table~\ref{tab2}. }
\label{f2}
\end{figure}

\subsection{X-ray Dominated Regions}

  Another concern in the $T_{\rm e}$-method analysis is the possible effect of 
 X-ray Dominated Regions (XDRs) on the emission-line spectra of AGNs. XDRs consist of 
a molecular region  mostly heated by direct photoionisation of the gas (primarily through the X-rays, 
which can penetrate deep into the cloud without dissociating molecules)
that can have an important contribution  
to the observed flux of hydrogen lines as well as of other lines (mainly in the infrared) observed in AGNs (e.g. \citealt{maloney96, meijerink05}).
The $T_{\rm e}$-method formalism (see Sect.~\ref{temeth}) assumes that most the flux of the emission-lines arise within  
the Str\"omgren sphere and the existence of additional flux from molecular/neutral gas introduces uncertainties on the abundances derived
by this method.
However, photoionization models simulations by \citet{ferland13} showed that the H$\beta$ flux emitted by XDRs is weaker by a factor of about $10^{3}$ than the one emitted by the Str\"omgren  sphere of an AGN.
This result indicates that the XDR effects on the abundance determinations based on the $T_{\rm e}$-method
is actually negligible.
 
\subsection{Gas shock}

A different approach is considering the presence of shocks caused by 
supersonic turbulence, jets, and/or winds as a secondary ionization source \citep{zhang13}. 
This extra  ionization and heating of the
gas  \citep{dopita95, dopita96} drives to high temperatures which in turn leads to derive unrealistically low abundance values through the $T_{\rm e}$-method.
If shocks have really an important contribution to the
ionization/heating of  AGNs, some correlation between temperatute and  shock indicators should be derived.
 A good tracer of the presence of shocks is, for instance, the FWHM of emission lines (e.g. \citealt{contini12}).
In order to test this scenario, in Fig.~\ref{f2},  the observational FWHM values 
for  the \citealt{dors17} sample versus
the difference between
the calculated and predicted  $t_{3}$ values, refered as ${\rm D}(t_{3})$, are shown.
It was possible to obtain FWHM values  for 17/26
objects listed in Table~\ref{tab2}.
As it can be noted, there is no correlation between  FWHM and ${\rm D}(t_{3})$: 
 the Pearson correlation coefficient is $-$0.012.   
  However, low shock velocities ($v \: \la \: 400 \: \rm km \:s^{-1}$) 
have been proposed to be in NLR of Seyfert 2s \citep{contini17a}. Moreover,  \citet{dopita95} found that
moderately low-velocity shocks ($v\: \sim 200\: \rm km \:s^{-1}$) can  produce a very 
large  [\ion{O}{iii}]$\lambda$4363/$\lambda$5007  flux ratio. Based on the previous analysis, 
we suggest that  is unlikely   
shocks are the main cause of the $t_{3}$ discrepancy, although
it is not possible to exclude them as the 
origin of part metallicity discrepancy focused in this work.

\subsection{Electron temperature fluctuation}

 Another possible cause of the $t_{3}$ discrepancy could be the presence of electron temperature fluctuations
in the gas phase of NLRs which should be more significant than those in \ion{H}{ii}
regions.  Spatially studies of  nearby \ion{H}{ii}
regions do not have derived  sufficient level of electron temperature fluctuation necessary to conciliate abundances based
on $T_{\rm e}$-method with those obtained using metal recombination lines or derived from
photoionization models (see, for example, \citealt{krabbe02, tsamis03, rubin03, esteban04, oliveira08}, among others). 
Regarding AGNs, spatially resolved abundance studies  are  seldomly found in the literature and
 few studies carried out in this direction have found a small temperature variations along
the radius of NLRs (\citealt{revalski18a, revalski18b}). Observations using the future class of giant telescopes
(e.g. Giant Magellanic Telescope) and data obtained with the James Webb Telescope could
reveal clumps of distinct temperature in NLRs and clarify the problem of abundance discrepancy in NLRs.

\begin{figure}
\centering
\includegraphics[angle=-90,width=1\columnwidth]{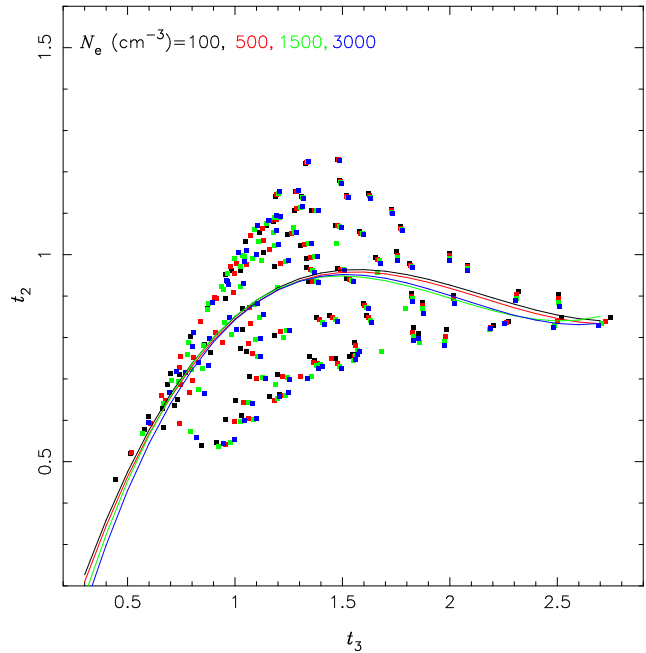}
\caption{Relation between $t_{2}$ and $t_{3}$ temperatures. Points represent
results of the photoionization model grid described in Sect.~\ref{model}.
Curves represent  the fittings (see Eq.~\ref{t2t3new}) to the 
 photoionization model results considering different electron density values, as indicated.}
\label{f3aa}
\end{figure}

\begin{figure}
\centering
\includegraphics[angle=-90,width=1\columnwidth]{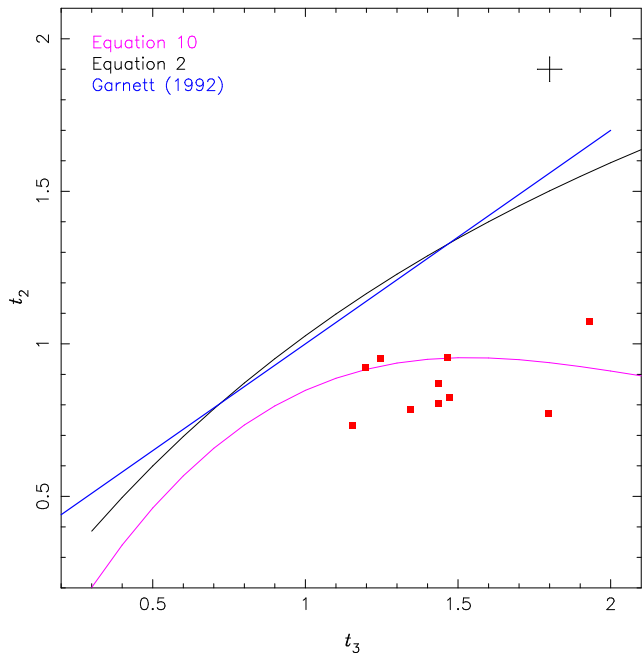}
\caption{Relation between $t_{2}$ and $t_{3}$ temperatures.
 Pink curve represents  the Eq.~\ref{t2t3new} with the fitted 
 coeficients considering all the assumed electron densities ($N_{\rm e}$=100, 500, 1500 and 3000 $\rm cm^{-3}$).  
Blue and black curves represent temperature relations for \ion{H}{ii}
regions predicted by photoionization models built by \citet[][Eq.~\ref{garn92}]{garnett92} 
and by \citet[][Eq.~\ref{eqt2}]{enrique14}, respectively.
 Points represent direct estimations of $t_{2}$ (calculated from $t_{e}$([\ion{N}{ii}]) and using the Eq.~\ref{eqt9a}) and 
 $t_{3}$  for some objects in our sample.
 Error bars represent the uncertainty ($\approx 800$ K) in the direct estimations of electron temperature (e.g. 
 \citealt{kennicutt03, hagele08}).}
\label{f3}
\end{figure}
 
\begin{figure}
\centering
\includegraphics[angle=-90,width=1\columnwidth]{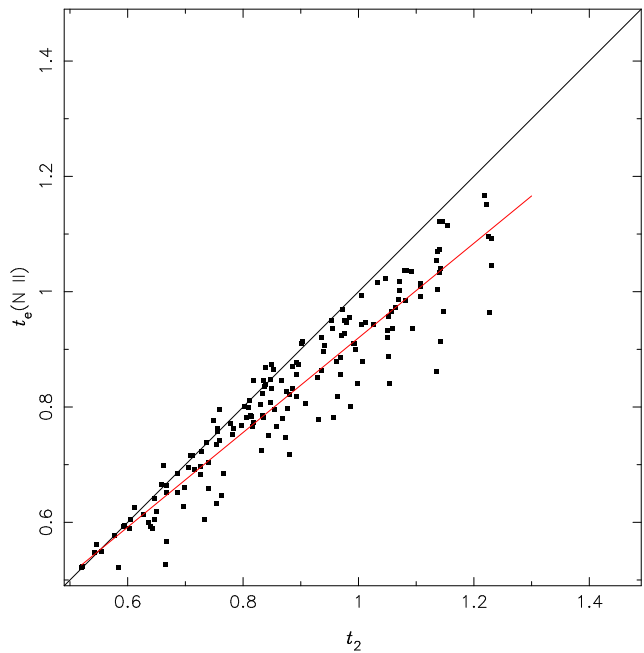}
\caption{$t_{\rm e}({\rm N\:II})$ versus $t_{2}$. Points represent
results of the photoionization model grid described in Sect.~\ref{model}. 
Black  line represents the equality between both estimations while
red line represents the fitting to the points given by the Eq.~\ref{eqt9a}.}
\label{f3aaa}
\end{figure} 

\subsection{$t_{2}-t_{3}$ relation}

Concerning the total oxygen abundance (O/H) discrepancy clearly noted in Fig.~\ref{f1}, its origin  can also be
 due to the use of the $t_{2}-t_{3}$ relation (Eq.~\ref{eqt2}) obtained through fitting values 
derived for \ion{H}{ii} regions, probably not representative for AGNs. 
In order to investigate this scenario, we used the results of the grid of
AGN photoionization models (see Sect.~\ref{model}) to obtain a new
$t_{2}-t_{3}$ relation, shown in Fig.~\ref{f3aa}. 
We also show in this figure the fitting to the expression: 
\begin{equation}
\label{t2t3new}
t_{2}=({\rm a} \times t_{3}^{3})+({\rm b} \times t_{3}^{2})+({\rm c} \times t_{3})+{\rm d},
\end{equation}
for the different $N_{\rm e}$ values.
It can be seen that  the resulting  $t_{2}$-$t_{3}$  fitting is independent on 
the $N_{\rm e}$ value adopted in the models. Therefore, we produced a new fitting
considering all points, not discriminating different $N_{\rm e}$ values, and found
for the coefficients of the expression above the values: 
$\rm a=0.17 \pm 0.04$, $\rm b=-1.07  \pm 0.22 $, $\rm c=2.07  \pm 0.32$ and $\rm d=-0.33  \pm 0.15$.  
It can be seen in Fig.~\ref{f3aa} that $t_{2}$ increases
with $t_{3}$ until $t_{3}\approx 1.5$,  remaining about constant for higher electron temperature
values. This result was also derived  by \citet{enrique09}, who  used the direct estimations of
$t_{3}$ for \ion{H}{ii} regions and the relation proposed by \citet{thurston96} to calculate $t_{2}$.

In Fig.~\ref{f3}, our derived $t_2$ -- $t_3$ relation, given by Eq.~\ref{t2t3new}, is compared
to the two relations found in the literature and estimated for \ion{H}{ii} regions using photoionization models.
 One of them is the relation proposed by \citet{garnett92}: 
\begin{equation}
\label{garn92}
t_{2}=\,0.7\,t_{3}+0.3
\end{equation}
and the other is the one by \citet[][see Eq.~\ref{eqt2}]{enrique14}.

We also show in Fig.~\ref{f3} the $t_{2}$ values calculated for some objects in our sample through the observational intensities
of the  $R_{N2}$=[\ion{N}{ii}]$\lambda$6548+$\lambda$6584/$\lambda$5755 
emission-line  ratio (intensities compiled from the same works than 
the other observational data, see Sect.~\ref{obs}).  The $t_{2}$ values
were obtained adopting the following procedure.
Firstly, we calculed the temperature for the $\rm N^{+}$ ion  using the relation
by \citet{hagele08}:
\begin{equation}
\label{eqt9}
 t_{\rm e}({\rm N\:II})=0.537+0.000\:253 \times R_{\rm N2}+\frac{42.13}{R_{\rm N2}}.
\end{equation}
 The critical density for the lines  envolved in the $R_{N2}$ ratio is $\rm \sim 10^{5} \: cm^{-3}$ \citep{appenzeller88},
a value lower than the one derived in NLRs of Seyfert~2 ($N_{\rm e} \: \la \: 1000 \: \rm  cm^{-3}$). Thus,
the Eq.~\ref{eqt9} is valid for the present study. 
In general, in abundance studies of \ion{H}{ii} regions  is  assumed $t_{2}$=$t_{\rm e}({\rm N\:II})$, as is commonly used  
when the [\ion{O}{ii}$]\lambda$7325 auroral emission-line sensitive 
to the temperature and the strong [\ion{O}{ii}$]\lambda$3727  line are not available  (see e.g.\ \citealt{kennicutt03, hagele08}).
 However,  this equality can not be  valid for AGNs. In order to verify this, in  Fig.~\ref{f3aaa},
we show the predictions of our grid of models for $t_{\rm e}({\rm N\:II})$ versus $t_{2}$. 
One can see a considerable difference between the temperatures, mainly for $t_{2} \: \ga \: 1$. In view of this, we used the model 
predictions   and derived the relation
\begin{equation}
\label{eqt9a}
t_{2}= 0.82  \times   t_{\rm e}({\rm N\:II}) + 0.1.
\end{equation}
%erro a +/- 0.02138
 %erro b +/- +/- 0.01907 
It is worth noting that the relations derived for \ion{H}{ii} regions are not representative for AGNs, producing, for a given $t_{3}$ value,
higher $t_{2}$ temperatures than those predicted by AGN models and, consequently, they lead to lower $\rm O^{+}/H^{+}$ values when 
derived through the $T_{\rm e}$-method.  The difference between the $t_{2}$-$t_{3}$ relations 
shown in Fig.~\ref{f3} is mainly due to the harder and distinct SED of AGNs leads to a higher input of
energy per photoionization, resulting in a different electron temperature structure than that in \ion{H}{ii} regions.
We can also see in this figure that the objects for which it was possible to directly derive the [\ion{O}{iii}] and [\ion{O}{ii}]
electron temperatures using the observational data  and the Eq.~\ref{eqt9a}
seem to follow the theoretical relation derived for AGNs.
Although direct determinations of $t_{3}$ by itself do not coincide with the values predicted by the 
models for AGNs, as previously  stated, surprisingly 
the relation between the observational and theoretical $t_{2}-t_{3}$ relations seem to be in agreement.
The same effects that we mentioned  may be producing the temperature problem could also be affecting
the observational [\ion{N}{ii}] temperature  (and consequently $t_{2}$) producing a kind of compensation that leads to the observed agreement.
However, the amount of observational data and its observed dispersion
do not allow us to infer a conclusive result.

\begin{figure}
\centering
\includegraphics[angle=-90,width=1\columnwidth]{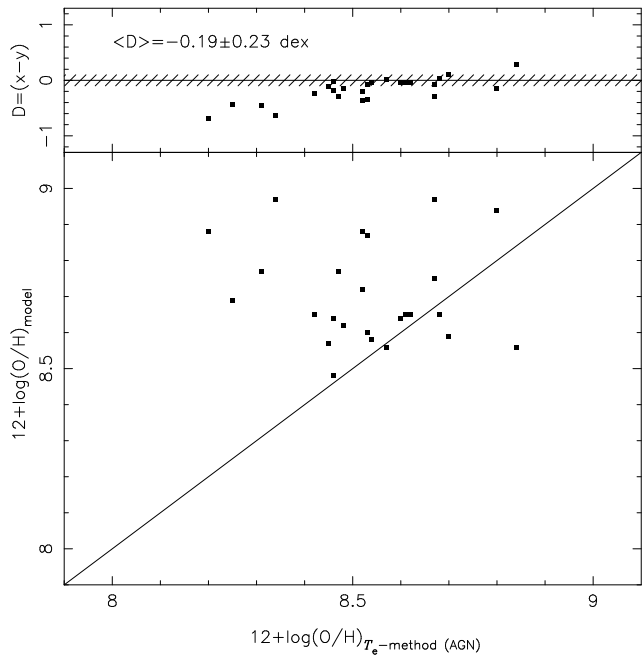}
\caption{Botton panel: Comparison between O/H values for the sample of objects (listed in Table~\ref{tab1})
obtained from detailed photoionization model (see Table~\ref{tab2}) versus the ones
computed by using the $T_{e}$-method   (see Sect.~\ref{temeth}) but assuming the
$t_{2}-t_{3}$-relation derived for AGNs (Eq.~\ref{t2t3new}). 
Solid line represents the equality between the estimations. Top panel:
Difference between these estimations (D=x-y). Solid line represents the null
difference between the estimations while   hatched area  represents
the uncertainty of 0.1 dex in abundances derived through $T_{\rm e}$-method (see \citealt{kennicutt03, hagele08}). 
The average difference is shown.}
\label{foxynew}
\end{figure}

In Fig.~\ref{foxynew}, we compare  the O/H abundance estimations for  the objects in the \citet{dors17}
sample 
computed by using the $T_{e}$-method expressions (see Sect.~\ref{temeth}) but assuming the
$t_{2}-t_{3}$ relation derived for AGNs, i.e.\ Eq.~\ref{t2t3new},  with those predicted
by the detailed photoionization models.
It can be seen that the difference between the estimations is
only significant for the low metallicity regime. Hence, the use of these
$t_{2}-t_{3}$ relation derived for AGNs and ICF(O) reduces the discrepancies
between the total oxygen abundance estimations.
Nevertheless, in the upper panel of Fig.~\ref{foxynew} there seem still to be a trend.
The average difference of about $-0.2$ dex between these two O/H estimations
is slightly higher (by about 0.1 dex) than that  found for \ion{H}{ii} regions by \citet{dors11} and \citet{enrique14}, who compared O/H estimations
for SFRs estimated by using the $T_{\rm e}$-method with those derived by 
photoionization models results. 

\begin{figure}
\centering
\includegraphics[angle=-90,width=1\columnwidth]{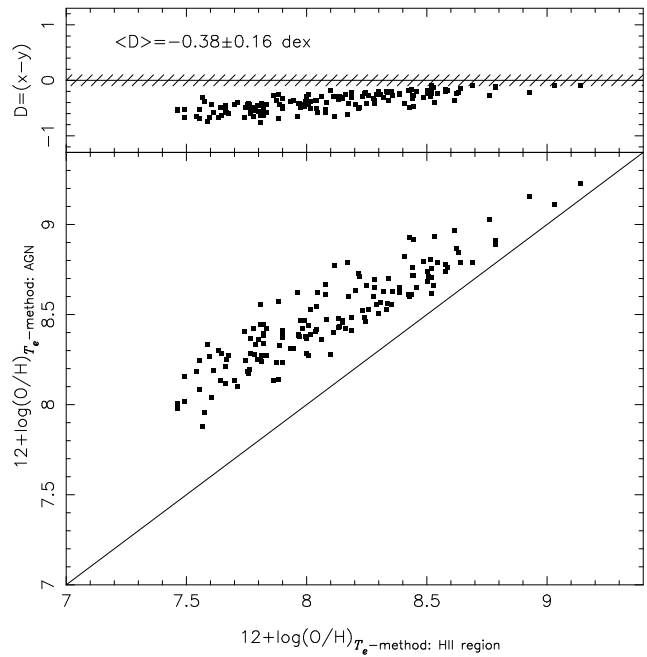}
\caption{Bottom panel: Comparison between  the total (O/H) abundances computed by the  
$T_{\rm e}$-method described in Sect.~\ref{temeth} with those computed by using also the $T_{\rm e}$-method
but assuming the $t_{2}-t_{3}$ relation derived for AGNs (Eq.~\ref{t2t3new}).
Points represent estimations for confirmed Seyfert 2s (redshift $z \la 0.4$) compiled from the SDSS  by \citet{dors20}. Solid line represents the equality between the estimations.
Top panel: Differences between these estimations (D=x-y) are shown. Solid line represents the null
difference between the estimations while   hatched area  represents
the uncertainty of 0.1 dex in abundances derived through the $T_{\rm e}$-method (see e.g.\ \citealt{kennicutt03, hagele08}).}
\label{f8}
\end{figure} 
 
We perform an additional analysis in order to compare O/H abundance estimations calculated  by using
the   $T_{\rm e}$-method,  assuming  the different $t_{2}$-$t_{3}$ relations (Eqs.~\ref{eqt2} and \ref{t2t3new})
as well as  assuming  calibrations between the metallicity (or O/H) with
strong emission-lines.  In view of this, the 
 \citet{dors20}  sample was used in this analysis.
Unfortunately, for the objects in this sample it is not possible
to calculate the ICF(O)  because the \ion{He}{ii}$\lambda$4686 emission line,
necessary to estimate $\rm N(He^{2+})$, was not measured.  Therefore, for all of these SDSS objects, we considered the ICF(O) to be equal
to 1.20,   an average value obtained from the ICFs  (obtained from Eq.~\ref{icfox})  listed in Table~\ref{tab2}. This assumed value
translates  into an oxygen abundance correction of about 0.1 dex. Regarding the calibration between O/H abundance and strong emission-lines, the first \citet{thaisa98}
theoretical calibration between  the oxygen abundance and  [N\,{\sc ii}]$\lambda$$\lambda$6548,6584/H$\alpha$ and  [O\,{\sc iii}]$\lambda$$\lambda$4959,5007/H$\beta$ 
line ratios is considered.
In \citet{dors20} a complete comparison between Seyfert~2  O/H abundances computed  through most of the methods available in the literature was carried out and it 
will not be repeated here.

\begin{figure}
\centering
\includegraphics[angle=-90,width=1\columnwidth]{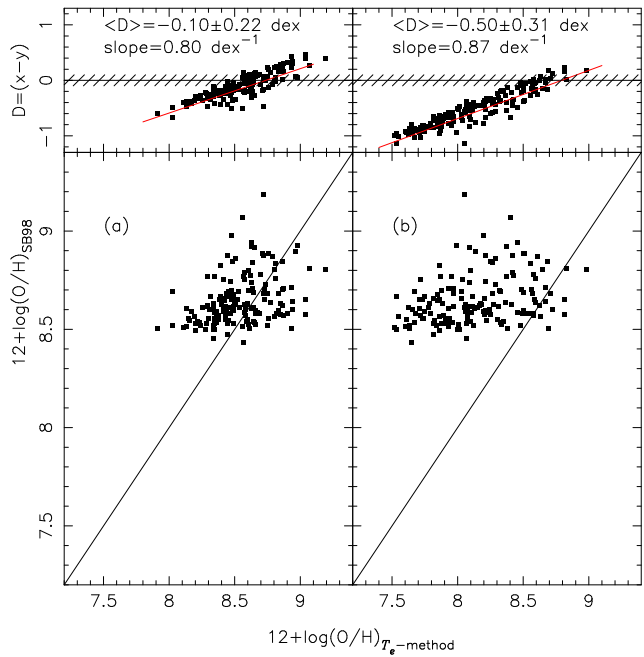}
\caption{Lower panels: total oxygen abundances estimated, for the objects in our SDSS sample,  using the first \citet{thaisa98}
theoretical calibration versus the ones calculated using the $T_{\rm e}$-method:  
in panel (a) the new formalism applying the $t_{2}-t_{3}$-relation derived for AGNs (Eq.~\ref{t2t3new}); in panel (b) the classical formalism applying the $t_{2}-t_{3}$-relation derived for \ion{H}{ii} regions (Eq.~\ref{eqt2}).
Upper panels: diferences between the estimations (D=x-y). }
\label{f10}
\end{figure}

In Fig.\ \ref{f8} we compare, for our SDSS sample, the O/H estimations obtained through 
the $T_{\rm e}$-method formalism developed for \ion{H}{ii}-regions and that developed for AGNs in 
the present work. In spite of the fact that we are not able to estimate the ICF(O) and have to apply a constant 
correction, we can see a systematic difference between the results obtained through these two methods 
greater than the 0.1 dex added by using the ICF(O) constant correction. The values estimated using the 
new formalism for AGNs are, in average, about 0.4 dex higher than those obtained through the formalism 
for \ion{H}{ii}-regions. As previously, the higher diferences are obtained for the lower metallicity values.
In Fig.~\ref{f10}, the results for estimations based on the first \citet{thaisa98} calibration
versus the ones calculated by using the $T_{\rm e}$-method for AGNs and for \ion{H}{ii} regions are shown. 
It can be seen that the  new formalism for the $T_{\rm e}$-method, i.e.\ applying the $t_{2}-t_{3}$-relation derived for AGNs,
reduces the discrepancy between the O/H values by about 0.40 dex when compared to those obtained by using 
the classical formalism for the $T_{\rm e}$-method, i.e. applying the $t_{2}-t_{3}$-relation derived for \ion{H}{ii} regions (Eq.~\ref{eqt2}).
It is worth mentioning that a systematic difference is still seen even though the point 
distribution has a lower scatter and is more tight around the one-to-one line. 
 
\section{Conclusions}
\label{conc}

We compiled from the literature narrow optical emission-line intensities of
26 Seyfert 2 AGNs in order to investigate the oxygen abundance (O/H) discrepancy
arising when compared the estimations by the classical
$T_{\rm e}$-method and by using detailed photoionization models.
 We found that the average  O/H discrepancy
($\sim 0.5$ dex) between the two methods is mainly due to the innapropriate 
use of the relation between  the tempetarure of the low ($t_{2}$) and high ($t_{3}$) ionization 
zones derived   for \ion{H}{ii} regions and generally used in the $T_{\rm e}$-method. 
Using results of a grid of photionization models, we derived an expression 
for the $t_{2}$-$t_{3}$ relation  which must be taken into account  in O/H estimations derived through the $T_{\rm e}$-method in chemical abundance studies of Seyfert~2 nuclei.
On the other hand, we use a second, more extensive, sample compiled from the SDSS database 
 to produce an additional comparison between O/H estimations obtained
  via the $T_{\rm e}$-method formalisms and also via a strong emission-line theoretical calibration. 
We found that the new formalism of the 
$T_{\rm e}$-method for AGNs produces O/H abundance higher by about 0.4 dex than
the ones obtained assuming the standard  equations  derived for \ion{H}{ii} regions.
Finally, we showed that the new formalism for the $T_{\rm e}$-method reduces by about 0.4 dex the O/H discrepancies
found when O/H abundances obtained from strong emission-line calibrations are compared to direct estimations.

\section*{Acknowledgments}

We are grateful to the anonymous referee for his/her very useful
comments and suggestions that helped us to clarify and improve
this work. OLD and ACK are grateful to FAPESP and CNPq.  
MA is grateful to CAPES. MVC and GFH are
grateful to CONICET. OLD thanks of kindly 
hospitality of the Kavli Institute for Cosmology staff where
part of this work was developed.
RM acknowledges ERC Advanced Grant 695671 ``QUENCH``  and support by the Science and Technology Facilities Council (STFC).
 
% =============================================================================================================

\label{lastpage}


\begin{thebibliography}{99}
\bibitem[Abazajian et al.(2009)]{abazajian09} Abazajian K.~N. et al., 2009, ApJS, 182, 543
\bibitem[Alende Prieto, Lambert \& Asplund(2001)]{allendeprieto01} Alende Prieto C., Lambert D.~L., Asplund M., 2001, ApJ, 556, L63
\bibitem[Allen et al.(2008)]{allen08} Allen M.~G., Groves B.~A, Dopita M.~A., Sutherland R.~S., Kewley L.~J., 2008, ApjSS, 178, 20 
\bibitem[Aller(1954)]{aller54} Aller L.~H., 1954, ApJ, 120, 401
\bibitem[Aller \& Liller(1959)]{aller59} Aller L.~H., \& Liller W., 1959, ApJ, 130, 45
\bibitem[Alloin et al.(1992)]{alloin92} Alloin  D., Bica  E., Bonatto  C., Prugniel P., 1992, A\&A, 266, 117
\bibitem[Appenzeller \& Oestreicher(1988)]{appenzeller88} Appenzeller I., \& Oestreicher R., 1988, AJ, 95, 45
\bibitem[Baron \& Netzer(2019)]{baron19} Baron D., \& Netzer H., 2019, MNRAS, 486, 4290
\bibitem[Baskin \& Laor(2005)]{baskin05} Baskin A., \&  Laor A., 2005, MNRAS, 358, 1043
\bibitem[Berg et al.(2020)]{berg20} Berg D.~A., Pogge R.~W.,  Skillman E.~D., Croxall K.~V., Moustakas J.~R., Noah S.~J.; Sun J., 2020, ApJ, 893, 96
\bibitem[Binette et al.(2012)]{binette12} Binette, L. et al., 2012, A\&A, 547, 29
\bibitem[Bowen \& Wyse(1939)]{bowen39} Bowen I.~S., \& Wyse A.~B., 1939, Lick Obs. Bull. 19, 1
\bibitem[Bresolin et al.(2004)]{bresolin04} Bresolin F., Garnett D.~R., Kennicutt R.~C.,  2004, ApJ, 615, 228
\bibitem[Castro et al.(2017)]{castro17} Castro C.~S., Dors O.~L., Cardaci M.~V., H{\"a}gele G.~F., 2017, MNRAS, 467, 1507
\bibitem[Carvalho et al.(2020)]{sarita20} Carvalho S.~P.  et al., 2020, MNRAS, 492, 5675
\bibitem[Cohen(1983)]{cohen83} Cohen R.~D., 1983, ApJ, 273, 489
\bibitem[Congiu et al.(2017)]{congiu17} Congiu E. et al., 2017, MNRAS, 471, 562
\bibitem[Contini(2017a)]{contini17a} Contini M., 2017a, MNRAS, 469, 3125
\bibitem[Contini(2017b)]{contini17b} Contini M., 2017b, MNRAS, 466, 2787
\bibitem[Contini et al.(2012)]{contini12}  Contini M., 2012, MNRAS, 425, 1205
\bibitem[Crenshaw \& Kraemer(2005)]{crenshaw05} Crenshaw D.~M., \& Kraemer S.~B., 2005, ApJ, 625, 680
\bibitem[Croxall et al.(2016)]{croxall16} Croxall K.~V., Pogge R.~W., Berg D.~A., Skillman E.~D., Moustakas J., 2016, ApJ, 830, 4
\bibitem[Cruz-Gonzalez et al.(1991)]{cruz91} Cruz-Gonzalez I., Guichard J., Serrano A., Carrasco L., 1991, PASP, 103, 888
\bibitem[Davies et al.(2020)]{davies20} Davies R. et al., 2020, arXiv:2003.06153
\bibitem[Dietrich et al.(2003)]{dietrich03} Dietrich M. et al., 2003, ApJ, 589, 722
\bibitem[Dopita \& Sutherland(1995)]{dopita95} Dopita M.~A., \& Sutherland R.~S., 1995, ApJ, 455, 468
\bibitem[Dopita et al.(1996)]{dopita96} Dopita M.~A., Groves B., Sutherland R.~S., 1996, ApJ, 102, 161
\bibitem[Dors \& Copetti(2005)]{dors05} Dors O.~L., Copetti M.~V.~F., 2005, A\&A, 437, 837
\bibitem[Dors et al.(2011)]{dors11} Dors O.~L., Krabbe \^A.~C., H\"agele G.~F., P\'erez-Montero E., MNRAS, 415, 3616
\bibitem[Dors et al.(2012)]{dors12} Dors, O.~L., Riffel R.~A., Cardaci, M.~V., H\"agele G.~F., Krabbe A.~C.; P\'erez-Montero E.,  Rodrigues I., 2012, MNRAS, 422, 252
\bibitem[Dors et al.(2013)]{dors13} Dors O.~L. et al., 2013, MNRAS, 432, 2512
\bibitem[Dors et al.(2014)]{dors14} Dors O.~L., Cardaci M.~V., H\"agele G.~F.,  Krabbe \^A.~C., 2014, MNRAS, 443, 1291
\bibitem[Dors et al.(2015)]{dors15} Dors O.~L., Cardaci M.~V., H\"agele G.~F., Rodrigues I., Grebel E.~K., Pilyugin, L.~S., Freitas-Lemes, P., Krabbe \^A.~C., 2015, MNRAS, 453, 4102
\bibitem[Dors et al.(2016)]{dors16} Dors O.~L., P\'erez-Montero E., H\"agele G.~F., Cardaci M.~V., Krabbe A.~C., MNRAS, 2016, 456, 4407
\bibitem[Dors et al.(2017)]{dors17} Dors O.~L., Arellano-C\'ordoba K.~Z., Cardaci M.~V., H\"agele G.~F., 2017, 468, L113
\bibitem[Dors et al.(2018)]{dors18} Dors O.~L., Agarwal B., H\"agele G.~F., Cardaci M.~V., Rydberg C., Riffel R.~A., Oliveira A.~S., Krabbe A.~C., 2018, MNRAS, 479, 2294
\bibitem[Dors et al.(2019)]{dors19} Dors O.~L., Monteiro A.~F, Cardaci M.~V., H\"agele G.~F.,  Krabbe \^A.~C., 2019, MNRAS, 486, 5853
\bibitem[Dors et al.(2020)]{dors20} Dors O.~L.  et al., 2020, MNRAS, 492, 468
\bibitem[Esteban et al.(2020)]{esteban19} Esteban C., Bresolin F., Garc{\'\i}a-Rojas J., Toribio San Cipriano L., 2020, MNRAS, 491,2137
\bibitem[Esteban et al.(2004)]{esteban04} Esteban C., Peimbert M.; Gar{\'\i}a-Rojas J., Ruiz M.~T., Peimbert A., Rodr{\'\i}guez M., 2004, MNRAS, 355, 229
\bibitem[Feltre, Charlot \& Gutkin(2016)]{feltre16} Feltre  A., Charlot S., Gutkin J., 2016, MNRAS, 456, 3354
\bibitem[Ferland \& Netzer(1983)]{ferland83} Ferland G.~J.,  \& Netzer H., 1983, ApJ, 264, 105
\bibitem[Ferland \& Osterbrock(1986)]{ferland86} Ferland G.~J., \& Osterbrock D.~E., 1986, ApJ, 300, 658
\bibitem[Ferguson et al.(1997)]{ferguson97} Ferguson J.~W., Korista K.~T.,  Baldwin J.~A., Ferland G.~J., 1997, ApJ, 487, 122
\bibitem[Fern\'andez et al.(2018)]{fernandez18} Fern\'andez V., Terlevich E., D\`{\i}az A.~I.; Terlevich R.,  Rosales-Ortega F.~F., 2018, MNRAS, 478, 5301
\bibitem[Ferland et al.(2013)]{ferland13} Ferland G.~J., 2013, Rev. Mex. Astron. Astrofis., 49, 137
\bibitem[Freitas et al.(2018)]{izabel18} Freitas I.~C. et al., 2018, MNRAS, 476, 2760
\bibitem[Garnett et al.(1997)]{garnett97} Garnett D.~R., Shields G.~A., Skillman E.~D., Sagan S.~P., Dufour R.~J.,  1997, ApJ, 489, 63
\bibitem[Garnett(1992)]{garnett92} Garnett D.~R., 1992, AJ, 103, 1330
\bibitem[Gburek et al.(2019)]{gburek19}  Gburek  T., Siana B.,  Alavi A., Emami N. et al., 2019, ApJ, 887, 168
\bibitem[Goodrich \& Osterbrock(1983)]{goodrich83} Goodrich R.~W., \& Osterbrock D., 1983, ApJ, 269, 416
\bibitem[Guo et al.(2020)]{guo20} Guo Y. et al., 2020, arXiv:2001.05473
\bibitem[Groves et al.(2006)]{groves06} Groves B.~A., Heckman T.~M., Kauffmann G., 2006, MNRAS, 371, 1559
\bibitem[H{\"a}gele et al.(2006)]{hagele06} H{\"a}gele, G.~F. et al., 2006, MNRAS, 372, 293
\bibitem[H{\"a}gele et al.(2008)]{hagele08} H{\"a}gele G.~F. et al.,  2008, MNRAS, 383, 209
\bibitem[H{\"a}gele et al.(2011)]{hagele11} H{\"a}gele, G.~F.  et al., 2011, MNRAS, 414, 272
\bibitem[H{\"a}gele et al.(2012)]{hagele12} H{\"a}gele, G.~F.  et al., 2012, MNRAS, 422, 3475
\bibitem[Heckman \& Balick(1979)]{heckman79} Heckman T.~M., \& Balick B., 1979, A\&A, 79, 350
\bibitem[Holt et al.(2011)]{holt11} Holt J., Tadhunter C., Morganti R., Emonts B., 2011, MNRAS, 410, 1527
\bibitem[Izotov et al.(1994)]{izotov94}Izotov Y.~I., Thuan T.~X., Lipovetsky V.~A., 1994, ApJ, 435, 647
\bibitem[Izotov et al.(2006)]{izotov06} Izotov Y.~I., Stasi\'nska G., Meynet G., Guseva N.~G.,  Thuan T.~X., 2006, A\&A, 448, 955
\bibitem[Izotov \& Thuan(2008)]{izotov08} Izotov Y.~I., \& Thuan T.~X., 2008, ApJ, 687, 133
\bibitem[Jensen et al.(1976)]{jensen76} Jensen E.~B., Strom K.~M., Strom S.~E., 1976, ApJ, 209, 748
\bibitem[Kakkad et al.(2018)]{kakkad18} Kakkad D. et al., 2018, A\&A, 618, 6
\bibitem[Kennicutt et al.(1989)]{kennicutt89} Kennicutt, R.~C., Keel, W.~C., \& Blaha, C.~A.\ 1989, AJ, 97, 1022
\bibitem[Kennicutt et al.(2003)]{kennicutt03} Kennicutt R.~C., Bresolin F., Garnett D. R., 2003, ApJ, 591, 801
\bibitem[Kewley et al.(2001)]{kewley01} Kewley L.~J., Dopita M.~A., Sutherland R.~S., Heisler C.~A., Trevena J., 2001, ApJ, 556, 121
\bibitem[Kewley \& Dopita(2002)]{kewley02} Kewley L.~J., \&  Dopita M.~A., 2002, ApJS, 142, 35
\bibitem[Kewley \& Ellison(2008)]{kewley08} Kewley L.~J., \&  Ellison S.~L., 2008, ApJ, 681, 1183
\bibitem[Komossa \& Schulz(1997)]{komossa97} Komossa S., \& Schulz H., 1997, A\&A, 323, 31
\bibitem[Koski(1978)]{koski78} Koski A.~T., 1978, ApJ, 223, 56
\bibitem[Krabbe \& Copetti(2002)]{krabbe02} Krabbe, A.~C., \&  Copetti M.~V.~C., 2002, A\&A, 387, 295
\bibitem[Krabbe et al.(2014)]{krabbe14} Krabbe, A.~C.  et al., 2014, MNRAS, 437, 1155
\bibitem[Kraemer et al.(1994)]{kraemer94} Kraemer S.~B., Wu C.-C., Crenshaw D.~M.,  Harrington J.~P., 1994, ApJ, 435, 171
\bibitem[Kraemer et al.(2011)]{kraemer11} Kraemer S.~B., Schmitt H.~R., Crenshaw D.~M., Melandez M., Turner T.~J.,  Guainazzi M., Mushotzky R.~F., 2011, ApJ, 727, 130
\bibitem[Lagos et al.(2009)]{lagos09} Lagos, P. et al., 2009, AJ, 137, 5068
\bibitem[Leitherer et al.(1999)]{leitherer99} Leitherer C. et al., 1999, ApJ, 123, 3
\bibitem[Lequeux et al.(1979)]{lequeux79} Lequeux J., Peimbert M., Rayo J.~F., Serrano A.,  Torres-Peimbert S. 1979, A\&A, 80, 155
\bibitem[Lin et al.(2017)]{lin17} Lin Z. et al., 2017, ApJ, 842, 97
\bibitem[L{\'o}pez-Hern{\'a}ndez et al.(2013)]{lopezh13} L{\'o}pez-Hern{\'a}ndez, J. et al., 2013, MNRAS, 430, 472
\bibitem[L\'opez-S\'anchez et al. (2007)]{angel07} L\'opez-S\'anchez. A., Esteban C., Garc\'{\i}a-Rojas J., Peimbert M.. Rodr\'{\i}guez M., 2007, ApJ, 656, 168
\bibitem[Maiolino et al.(2008)]{maiolino08} Maiolino R. et al., 2008, A\&A, 488, 463
\bibitem[Maiolino \& Mannucci (2019)]{maiolino19} Maiolino R., \& Mannucci F., 2019, A\&A Rev., 27, 3
\bibitem[Maloney et al.(1996)]{maloney96} Maloney P.~R., Hollenbach D.~J., Tielens A.~G.~G.~M., 1996, ApJ, 466, 561
\bibitem[Matsuoka et al.(2009)]{matsuoka09} Matsuoka K., Nagao T., Maiolino R., Marconi A., TaniguchiY., 2009, A\&A, 503, 721
\bibitem[Matsuoka et al.(2018)]{matsuoka18} Matsuoka, K. et al., 2018, A\&A, 616, L4 
\bibitem[Meijerink \& Spaans(2005)]{meijerink05} Meijerink R., \& Spaans, 2005,  A\&A 436, 397
\bibitem[Mingozzi et al.(2019)]{mingozzi19} Mingozzi M. et al., 2019, A\&A, 622, 146
\bibitem[Mignoli et al.(2019)]{mignoli19} Mignoli M. et al., 2019, A\&A, 626, 9
\bibitem[Moll\'a \& D\'{\i}az (2005)]{molla05} Moll\'a M., \& D\'{\i}az A.~I., 2005, MNRAS, 358, 521
\bibitem[Nagao et al. (2001)]{nagao01} Nagao T., Murayama T., Taniguchi Y., 2001, ApJ, 549, 155
\bibitem[Nagao et al.(2006)]{nagao06a} Nagao, T., Maiolino, R., Marconi, A., 2006, A\&A, 447, 863
\bibitem[Nakajima et al.(2018)]{nakajima18} Nakajima K.  et al., 2018, A\&A, 612, 94
\bibitem[Nicholls et al. (2012)]{nicholls12} Nicholls D.~C., Dopita M.~A., Sutherland R.~S., 2012, ApJ, 752, 148 
\bibitem[Oliveira et al.(2008)]{oliveira08} Oliveira V.~A., Copetti M.~V.~F., Krabbe A.~C., 2008, A\&A, 492, 463
\bibitem[Osterbrock \& Miller(1975)]{osterbrock75} Osterbrock D.~E., \& Miller J. S., 1975, ApJ , 197, 535
\bibitem[Osterbrock \& Ferland(2006)]{Osterbrock06} Osterbrock, D.~E., \& Ferland, G.~J.\ 2006, Astrophysics of gaseous nebulae and active galactic nuclei
\bibitem[Page(1936)]{page36} Page T., 1936, Nature, 138, 503
\bibitem[Pagel et al.\ (1979)]{pagel79} Pagel B. E. J., Edmunds M.~G., Blackwell D.~E., Chun M.~S., Smith G., 1979, MNRAS, 189, 95
\bibitem[P\'erez-Montero et al.(2019)]{enrique19} P\'erez-Montero E. et al.,  2019, MNRAS, 489, 2652  
\bibitem[P\'erez-Montero(2017)]{enrique17} P\'erez-Montero E., 2017, PASP, 129, 043001
\bibitem[P\'erez-Montero(2014)]{enrique14}  P\'erez-Montero  E., 2014, MNRAS, 441, 2663
\bibitem[P\'erez-Montero et al. (2010)]{enrique10}  P\'erez-Montero  E., Garc\'{\i}a-Benito R., H\"agele G.~F., D\'{\i}az A.~I., 2010, MNRAS, 404, 2037
\bibitem[P\'erez-Montero \& Contini(2009)]{enrique09}  P\'erez-Montero  E., \& Contini T., 2009, MNRA, 398, 949
\bibitem[P\'erez-Montero \& D\'{\i}az(2003)]{enrique03} P\'erez-Montero E., \& D\'{\i}az A.~I., 2003, MNRAS, 346, 105 
\bibitem[Peimbert(1967)]{peimbert67} Peimbert M., 1967, ApJ, 150, 825
\bibitem[Peimbert et al. (1978)]{peimbert78} Peimbert M., Torres-Peimbert S., Rayo J.~F., 1978, ApJ, 220, 516
\bibitem[Peimbert et al. (2017)]{peimbert17} Peimbert M.,  Peimbert A.,  Delgado-Inglada G., 2017, PASP, 129, 082001
\bibitem[Pilyugin et al.(2004)]{pilyugin04} Pilyugin L.~S., V\'{\i}lchez J.~M., Contini T., 2004, A\&A, 425, 849
\bibitem[Phillips et al.(1983)]{phillips83} Phillips M.~M., Charles P.~A., Baldwin J.~A., 1983, ApJ, 266, 485
\bibitem[Rela{\~n}o et al.(2010)]{relanio10} Rela{\~n}o M.   et al., 2010, MNRAS, 402, 1635
\bibitem[Revalski et al.(2018a)]{revalski18a} Revalski M. et al., 2018a, ApJ, 867, 88
\bibitem[Revalski et al.(2018b)]{revalski18b} Revalski M., Crenshaw D.~M., Kraemer S.~B.; Fischer  T.~C., Schmitt H.~R., Machuca C., 2018b, ApJ, 856, 46
\bibitem[Riffel et al.(2013)]{riffel13} Riffel R., Rodr\'{i}guez-Ardila A., Aleman I., Brotherton M.~S., Pastoriza M.~G., Bonatto C., Dors O.~L., 2013, MNRAS, 430, 2002
\bibitem[Rubin et al.(2003)]{rubin03} Rubin R.~H., Martin P.~G., Dufour R.~J., Ferland G.~J., Blagrave K.~P.~M., Liu X.-W., Nguyen J.~F., Baldwin J.~A., 2003, MNRAS, 340, 362
\bibitem[Sanders et al.(2016)]{sanders16} Sanders R.~L.  et al., 2016, ApJ, 825, L23
\bibitem[Sanders et al.(2020)]{sanders20} Sanders R.~L.  et al., 2020, MNRAS, 491, 1427
\bibitem[Schmitt et al.(1994)]{enrique94} Schmitt H.~R., Storchi-Bergmann T., Baldwin J.~A., 1994, ApJ, 423, 237
\bibitem[Shuder(1980)]{shuder80} Shuder J.~M., 1980, ApJ, 240, 32
\bibitem[Shuder \&  Osterbrock(1981)]{shuder81} Shuder J.~M., \&  Osterbrock D.~E., 1981, ApJ, 250, 55
\bibitem[Smith (1975)]{smith75} Smith H.~E., 1975, ApJ, 199, 591
\bibitem[Stasi\'nska(1984)]{grazina84} Stasi\'nska G., 1984, A\&A, 135, 341
\bibitem[Storchi-Bergmann et al.(1998)]{thaisa98} Storchi-Bergmann T.,  Schmitt H.~R.,  Calzetti D., Kinney A.~L., 1998, AJ, 115, 909 
\bibitem[Thomas et al.(2019)]{thomas19} Thomas A.~D., Kewley L.~J., Dopita M.~A., Groves B.~A., Hopkins A.~M., Sutherland R.~S., 2019, ApJ, 874, 100
\bibitem[Thurston, Edmunds \& Henry(1996)]{thurston96} Thurston  T.~R., Edmunds M.~G., Henry R.~B.~C., 1996, MNRAS, 293, 990 
\bibitem[Torres-Peimbert  \& Peimbert(1977)]{torrespeimber77} Torres-Peimbert S., \& Peimbert M., 1977, Rev. Mex. Astron. Astrofis., 2, 181
\bibitem[Torres-Peimbert et al.(1989)]{torrespeimber89} Torres-Peimbert S., Peimbert M., Fierro J., 1989, ApJ, 345, 186
\bibitem[Tsamis et al.(2003)]{tsamis03} Tsamis Y.~G., Barlow M.~J., Liu X.-W., Danziger I.~J., Storey P.~J., 2003, MNRAS, 338, 687 
\bibitem[van Hoof(1997)]{vanhoof97} van Hoof P.~A.~M. Photo-ionization studies of nebulae. PhD thesis,  Rijksuniversiteit Groningen, 1997
\bibitem[van Zee et al.(1998)]{vanzee98} van Zee L., Salzer J.~J., Haynes M.~P.., O'Donoghue  A.~A., Balonek T.~J.,  1998, AJ, 116, 2805
\bibitem[Vaona et al.(2012)]{vaona12} Vaona L., Ciroi S., Di Mille F., Cracco V., La Mura G., Rafanelli P., 2012, MNRAS, 427, 1266
\bibitem[Viegas(2002)]{viegas02} Viegas S.~M., 2002, Rev. Mex. Astron. Astrofis., 12, 219
\bibitem[Vila-Costas \& Edmunds(1992)]{vila92} Vila-Costas M.~B, \&  Edmunds M.~G., 1992, MNRAS, 259, 121
\bibitem[Wyse(1942)]{wyse42} Wyse A.~B., 1942, ApJ, 95, 356
\bibitem[Yates et al.(2012)]{yates12} Yates R.~M., Kauffmann G., Guo Q., 2012, MNRAS, 422, 215
\bibitem[York et al.(2000)]{york00} York D.~G. et al., 2000, ApJ, 120, 1579
\bibitem[Zaritsky et al.(1994)]{zaritsky94} Zaritsky, D., Kennicutt, R.~C.,  Huchra, J.~P., 1994, ApJ, 420, 87
\bibitem[Zhang et al.(2013)]{zhang13} Zhang Z.~T., Liang Y.~C.; Hammer F., 2013, MNRAS, 430, 2605
\bibitem[Zinchenko et al.(2019)]{igor19} Zinchenko I.~A., Dors O.~L., H\"agele G.~F., Cardaci M.~V., Krabbe A.~C., 2019, MNRAS, 483, 1901
\bibitem[Zurita \& Bresolin(2012)]{zurita12} Zurita A., \& Bresolin F.,  2012, MNRAS, 427, 1463
\end{thebibliography}
\end{document}